\documentclass[twocolumn,letterpaper,aps,prc,longbibliography,superscriptaddress,nofootinbib,floatfix]{revtex4-1}

\usepackage{graphicx}	

\usepackage{amsmath}	
\usepackage{eufrak}
\usepackage{multirow}
\usepackage{amssymb}
\usepackage{xspace}	

\newcommand{\snn}{\mbox{$\sqrt{s_{_{NN}}}$}\xspace}

\newcommand{\la}{\langle}
\newcommand{\ra}{\rangle}

\newcommand{\eps}{\varepsilon}
\newcommand{\mean}[1]{\la #1 \ra}

\newcommand{\vnt}{v_n\{2\}}
\newcommand{\vnf}{v_n\{4\}}
\newcommand{\vns}{v_n\{6\}}
\newcommand{\vne}{v_n\{8\}}
\newcommand{\cnt}{c_n\{2\}}
\newcommand{\cnf}{c_n\{4\}}
\newcommand{\crf}{c_3\{4\}}
\newcommand{\cnsix}{c_n\{6\}}
\newcommand{\cne}{c_n\{8\}}

\newcommand{\vtt}{v_2\{2\}}
\newcommand{\vtf}{v_2\{4\}}
\newcommand{\vts}{v_2\{6\}}
\newcommand{\vte}{v_2\{8\}}

\newcommand{\ctf}{c_2\{4\}}

\newcommand{\nfvtxt}{N_{\rm tracks}^{\rm FVTX}}
\newcommand{\sigman}{\sigma_{v_n}}
\newcommand{\sigmat}{\sigma_{v_2}}
\newcommand{\sigmar}{\sigma_{v_3}}
\newcommand{\sigmaon}{\sigma_{v_n}/\mean{v_n}}
\newcommand{\sigmaot}{\sigma_{v_2}/\mean{v_2}}
\newcommand{\sigmaor}{\sigma_{v_3}/\mean{v_3}}
\newcommand{\npart}{N_{\rm part}}
\newcommand{\chisq}{\chi^2/{\rm d.o.f.}}

\newcommand{\pp}{\mbox{$p$$+$$p$}\xspace}
\newcommand{\pau}{\mbox{$p$$+$Au}\xspace}

\newcommand{\auau}{\mbox{Au$+$Au}\xspace}
\newcommand{\pbpb}{\mbox{Pb$+$Pb}\xspace}

\newcommand{\geant}{{\sc geant}-4\xspace}
\newcommand{\ampt}{{\sc ampt}\xspace}

\begin{document}

\title{Multi-particle azimuthal correlations for extracting
event-by-event elliptic and triangular flow in Au$+$Au collisions at
$\sqrt{s_{_{NN}}}=200$ GeV}

\newcommand{\abilene}{Abilene Christian University, Abilene, Texas 79699, USA}
\newcommand{\augie}{Department of Physics, Augustana University, Sioux Falls, South Dakota 57197, USA}
\newcommand{\banaras}{Department of Physics, Banaras Hindu University, Varanasi 221005, India}
\newcommand{\barc}{Bhabha Atomic Research Centre, Bombay 400 085, India}
\newcommand{\baruch}{Baruch College, City University of New York, New York, New York, 10010 USA}
\newcommand{\bnlcoll}{Collider-Accelerator Department, Brookhaven National Laboratory, Upton, New York 11973-5000, USA}
\newcommand{\bnlphys}{Physics Department, Brookhaven National Laboratory, Upton, New York 11973-5000, USA}
\newcommand{\caucr}{University of California-Riverside, Riverside, California 92521, USA}
\newcommand{\charlesczech}{Charles University, Ovocn\'{y} trh 5, Praha 1, 116 36, Prague, Czech Republic}
\newcommand{\chonbuk}{Chonbuk National University, Jeonju, 561-756, Korea}
\newcommand{\ciae}{Science and Technology on Nuclear Data Laboratory, China Institute of Atomic Energy, Beijing 102413, People's Republic of China}
\newcommand{\cns}{Center for Nuclear Study, Graduate School of Science, University of Tokyo, 7-3-1 Hongo, Bunkyo, Tokyo 113-0033, Japan}
\newcommand{\colorado}{University of Colorado, Boulder, Colorado 80309, USA}
\newcommand{\columbia}{Columbia University, New York, New York 10027 and Nevis Laboratories, Irvington, New York 10533, USA}
\newcommand{\czechtech}{Czech Technical University, Zikova 4, 166 36 Prague 6, Czech Republic}
\newcommand{\debrecen}{Debrecen University, H-4010 Debrecen, Egyetem t{\'e}r 1, Hungary}
\newcommand{\elte}{ELTE, E{\"o}tv{\"o}s Lor{\'a}nd University, H-1117 Budapest, P{\'a}zm{\'a}ny P.~s.~1/A, Hungary}
\newcommand{\eszterhazy}{Eszterh\'azy K\'aroly University, K\'aroly R\'obert Campus, H-3200 Gy\"ongy\"os, M\'atrai \'ut 36, Hungary}
\newcommand{\ewha}{Ewha Womans University, Seoul 120-750, Korea}
\newcommand{\fsu}{Florida State University, Tallahassee, Florida 32306, USA}
\newcommand{\gsu}{Georgia State University, Atlanta, Georgia 30303, USA}
\newcommand{\hiroshima}{Hiroshima University, Kagamiyama, Higashi-Hiroshima 739-8526, Japan}
\newcommand{\howard}{Department of Physics and Astronomy, Howard University, Washington, DC 20059, USA}
\newcommand{\ihepprot}{IHEP Protvino, State Research Center of Russian Federation, Institute for High Energy Physics, Protvino, 142281, Russia}
\newcommand{\illuiuc}{University of Illinois at Urbana-Champaign, Urbana, Illinois 61801, USA}
\newcommand{\inrras}{Institute for Nuclear Research of the Russian Academy of Sciences, prospekt 60-letiya Oktyabrya 7a, Moscow 117312, Russia}
\newcommand{\instpasczech}{Institute of Physics, Academy of Sciences of the Czech Republic, Na Slovance 2, 182 21 Prague 8, Czech Republic}
\newcommand{\isu}{Iowa State University, Ames, Iowa 50011, USA}
\newcommand{\jaea}{Advanced Science Research Center, Japan Atomic Energy Agency, 2-4 Shirakata Shirane, Tokai-mura, Naka-gun, Ibaraki-ken 319-1195, Japan}
\newcommand{\jyvaskyla}{Helsinki Institute of Physics and University of Jyv{\"a}skyl{\"a}, P.O.Box 35, FI-40014 Jyv{\"a}skyl{\"a}, Finland}
\newcommand{\kek}{KEK, High Energy Accelerator Research Organization, Tsukuba, Ibaraki 305-0801, Japan}
\newcommand{\korea}{Korea University, Seoul 02841, Korea}
\newcommand{\kurchatov}{National Research Center ``Kurchatov Institute", Moscow, 123098 Russia}
\newcommand{\kyoto}{Kyoto University, Kyoto 606-8502, Japan}
\newcommand{\lawllnl}{Lawrence Livermore National Laboratory, Livermore, California 94550, USA}
\newcommand{\losalamos}{Los Alamos National Laboratory, Los Alamos, New Mexico 87545, USA}
\newcommand{\lund}{Department of Physics, Lund University, Box 118, SE-221 00 Lund, Sweden}
\newcommand{\lyon}{IPNL, CNRS/IN2P3, Univ Lyon, Université Lyon 1, F-69622, Villeurbanne, France}
\newcommand{\maryland}{University of Maryland, College Park, Maryland 20742, USA}
\newcommand{\mass}{Department of Physics, University of Massachusetts, Amherst, Massachusetts 01003-9337, USA}
\newcommand{\michigan}{Department of Physics, University of Michigan, Ann Arbor, Michigan 48109-1040, USA}
\newcommand{\muhlenberg}{Muhlenberg College, Allentown, Pennsylvania 18104-5586, USA}
\newcommand{\nara}{Nara Women's University, Kita-uoya Nishi-machi Nara 630-8506, Japan}
\newcommand{\natmephi}{National Research Nuclear University, MEPhI, Moscow Engineering Physics Institute, Moscow, 115409, Russia}
\newcommand{\newmex}{University of New Mexico, Albuquerque, New Mexico 87131, USA}
\newcommand{\nmsu}{New Mexico State University, Las Cruces, New Mexico 88003, USA}
\newcommand{\northcg}{Physics and Astronomy Department, University of North Carolina at Greensboro, Greensboro, North Carolina 27412, USA}
\newcommand{\ohio}{Department of Physics and Astronomy, Ohio University, Athens, Ohio 45701, USA}
\newcommand{\ornl}{Oak Ridge National Laboratory, Oak Ridge, Tennessee 37831, USA}
\newcommand{\orsay}{IPN-Orsay, Univ.~Paris-Sud, CNRS/IN2P3, Universit\'e Paris-Saclay, BP1, F-91406, Orsay, France}
\newcommand{\peking}{Peking University, Beijing 100871, People's Republic of China}
\newcommand{\pnpi}{PNPI, Petersburg Nuclear Physics Institute, Gatchina, Leningrad region, 188300, Russia}
\newcommand{\riken}{RIKEN Nishina Center for Accelerator-Based Science, Wako, Saitama 351-0198, Japan}
\newcommand{\rikjrbrc}{RIKEN BNL Research Center, Brookhaven National Laboratory, Upton, New York 11973-5000, USA}
\newcommand{\rikkyo}{Physics Department, Rikkyo University, 3-34-1 Nishi-Ikebukuro, Toshima, Tokyo 171-8501, Japan}
\newcommand{\saispbstu}{Saint Petersburg State Polytechnic University, St.~Petersburg, 195251 Russia}
\newcommand{\seoulnat}{Department of Physics and Astronomy, Seoul National University, Seoul 151-742, Korea}
\newcommand{\stonybrkc}{Chemistry Department, Stony Brook University, SUNY, Stony Brook, New York 11794-3400, USA}
\newcommand{\stonycrkp}{Department of Physics and Astronomy, Stony Brook University, SUNY, Stony Brook, New York 11794-3800, USA}
\newcommand{\tenn}{University of Tennessee, Knoxville, Tennessee 37996, USA}
\newcommand{\titech}{Department of Physics, Tokyo Institute of Technology, Oh-okayama, Meguro, Tokyo 152-8551, Japan}
\newcommand{\tsukuba}{Tomonaga Center for the History of the Universe, University of Tsukuba, Tsukuba, Ibaraki 305, Japan}
\newcommand{\vandy}{Vanderbilt University, Nashville, Tennessee 37235, USA}
\newcommand{\weizmann}{Weizmann Institute, Rehovot 76100, Israel}
\newcommand{\wigner}{Institute for Particle and Nuclear Physics, Wigner Research Centre for Physics, Hungarian Academy of Sciences (Wigner RCP, RMKI) H-1525 Budapest 114, POBox 49, Budapest, Hungary}
\newcommand{\yonsei}{Yonsei University, IPAP, Seoul 120-749, Korea}
\newcommand{\zagreb}{Department of Physics, Faculty of Science, University of Zagreb, Bijeni\v{c}ka c.~32 HR-10002 Zagreb, Croatia}
\affiliation{\abilene}
\affiliation{\augie}
\affiliation{\banaras}
\affiliation{\barc}
\affiliation{\baruch}
\affiliation{\bnlcoll}
\affiliation{\bnlphys}
\affiliation{\caucr}
\affiliation{\charlesczech}
\affiliation{\chonbuk}
\affiliation{\ciae}
\affiliation{\cns}
\affiliation{\colorado}
\affiliation{\columbia}
\affiliation{\czechtech}
\affiliation{\debrecen}
\affiliation{\elte}
\affiliation{\eszterhazy}
\affiliation{\ewha}
\affiliation{\fsu}
\affiliation{\gsu}
\affiliation{\hiroshima}
\affiliation{\howard}
\affiliation{\ihepprot}
\affiliation{\illuiuc}
\affiliation{\inrras}
\affiliation{\instpasczech}
\affiliation{\isu}
\affiliation{\jaea}
\affiliation{\jyvaskyla}
\affiliation{\kek}
\affiliation{\korea}
\affiliation{\kurchatov}
\affiliation{\kyoto}
\affiliation{\lawllnl}
\affiliation{\losalamos}
\affiliation{\lund}
\affiliation{\lyon}
\affiliation{\maryland}
\affiliation{\mass}
\affiliation{\michigan}
\affiliation{\muhlenberg}
\affiliation{\nara}
\affiliation{\natmephi}
\affiliation{\newmex}
\affiliation{\nmsu}
\affiliation{\northcg}
\affiliation{\ohio}
\affiliation{\ornl}
\affiliation{\orsay}
\affiliation{\peking}
\affiliation{\pnpi}
\affiliation{\riken}
\affiliation{\rikjrbrc}
\affiliation{\rikkyo}
\affiliation{\saispbstu}
\affiliation{\seoulnat}
\affiliation{\stonybrkc}
\affiliation{\stonycrkp}
\affiliation{\tenn}
\affiliation{\titech}
\affiliation{\tsukuba}
\affiliation{\vandy}
\affiliation{\weizmann}
\affiliation{\wigner}
\affiliation{\yonsei}
\affiliation{\zagreb}
\author{A.~Adare} \affiliation{\colorado} 
\author{C.~Aidala} \affiliation{\michigan} 
\author{N.N.~Ajitanand} \altaffiliation{Deceased} \affiliation{\stonybrkc} 
\author{Y.~Akiba} \email[PHENIX Spokesperson: ]{akiba@rcf.rhic.bnl.gov} \affiliation{\riken} \affiliation{\rikjrbrc} 
\author{M.~Alfred} \affiliation{\howard} 
\author{N.~Apadula} \affiliation{\isu} \affiliation{\stonycrkp} 
\author{H.~Asano} \affiliation{\kyoto} \affiliation{\riken} 
\author{B.~Azmoun} \affiliation{\bnlphys} 
\author{V.~Babintsev} \affiliation{\ihepprot} 
\author{A.~Bagoly} \affiliation{\elte} 
% ...
\collaboration{PHENIX Collaboration} \noaffiliation

\date{\today}


\begin{abstract}

We present measurements of elliptic and triangular azimuthal anisotropy
of charged particles detected at forward rapidity $1<|\eta|<3$ in
Au$+$Au collisions at $\sqrt{s_{_{NN}}}=200$ GeV, as a function of
centrality. The multiparticle cumulant technique is used to obtain the
elliptic flow coefficients $v_2\{2\}$, $v_2\{4\}$, $v_2\{6\}$, and
$v_2\{8\}$, and triangular flow coefficients $v_3\{2\}$ and $v_3\{4\}$.
Using the small-variance limit, we estimate the mean and variance of the
event-by-event $v_2$ distribution from $v_2\{2\}$ and $v_2\{4\}$.  In a
complementary analysis, we also use a folding procedure to study the
distributions of $v_2$ and $v_3$ directly, extracting both the mean and
variance.  Implications for initial geometrical fluctuations and their
translation into the final state momentum distributions are discussed.

\end{abstract}

\maketitle

\section{Introduction}
\label{sec:introduction}

Collisions of heavy nuclei at ultra-relativistic energies are believed
to create a state of matter called the strongly coupled quark-gluon
plasma, as first observed at the Relativistic Heavy Ion Collider
(RHIC)~\cite{Adcox:2004mh,Adams:2005dq,Back:2004je,Arsene:2004fa}. The
quark-gluon plasma evolves hydrodynamically as a nearly perfect liquid
as evinced by the wealth of experimental measurements and theoretical
predictions (or descriptions) of the azimuthal anisotropy of the
produced particles.~\cite{Romatschke:2017ejr}. Multi-particle
correlations are generally taken as strong evidence of hydrodynamical
flow, which necessarily affects most or all particles in the
event~\cite{Heinz:2013th}.  This is different from mimic correlations
(generically called nonflow) that are not related to the hydrodynamical
evolution and typically involve only a few particles.

Multi-particle correlations are also interesting because they have
different sensitivities to the underlying event-by-event fluctuations,
which can provide additional insights into the initial geometry and its
translation into final state particle
distributions~\cite{Voloshin:2007pc,Ollitrault:2009ie}.

Recently, experimental and theoretical efforts have been directed
towards measuring the fluctuations directly, using event-by-event
unfolding techniques. In principle, the multi-particle correlations and
unfolding techniques provide the same information about the underlying
fluctuations, though in practice with different
sensitivities~\cite{Jia:2014jca}.  The techniques used at the
Large Hadron Collider (LHC) are experimentally
very different and provide complementary
information~\cite{Aad:2013xma,Chatrchyan:2013kba}.

In this manuscript we present measurements of 2-, 4-, 6-, and 8-particle
correlations as well as event-by-event measurements of the azimuthal
anisotropy parameters corresponding to elliptic $v_2$ and triangular
$v_3$ flow. We estimate the relationship between the mean and variance
with both techniques and discuss the implications for understanding the
detailed shape of the $v_2$ and $v_3$ distributions. These measurements,
while the first of their kind at forward rapidity, are consistent with
previous measurements at midrapidity by STAR~\cite{Adams:2004bi} and
PHOBOS~\cite{Alver:2010rt}.

\section{Experimental Setup}
\label{sec:experiment}

In 2014, the PHENIX experiment~\cite{Adcox:2003zm} at RHIC collected nearly
$2\times10^{10}$ minimum bias (MB) events of Au$+$Au collisions at a
nucleon-nucleon center-of-mass energy $\snn$ = 200 GeV. The present
analysis makes use of a subset ($\approx10^9$ events) of the total 2014
data sample. The PHENIX beam-beam counters (BBC) are used for triggering
and centrality determination. The BBCs~\cite{Allen:2003zt} are located
$\pm$~144~cm from the nominal interaction point and cover the full
azimuth and $3.1<|\eta|<3.9$ in pseudorapidity. By convention, the north
side is forward rapidity ($\eta>0$) and the south side is backward
rapidity ($\eta<0$). Each BBC comprises an array of 64 phototubes with a
fused quartz \v{C}erenkov radiator on the front.  Charged particles
impinging on the radiator produce \v{C}erenkov light which is then
amplified and detected by the phototube. The PHENIX MB trigger for the
2014 data sample of Au$+$Au collisions at $\snn$ = 200 GeV was defined by
at least two phototubes in each side of the BBC having signal above
threshold and an online $z$-vertex within $\pm$~10~cm of the nominal
interaction point.  Additionally, PHENIX has a set of zero-degree
calorimeters (ZDC) that measure spectator neutrons from each incoming
nucleus~\cite{Allen:2003zt}.  We require a minimum energy in both ZDCs
to remove beam related background present at the highest luminosities.

The centrality definition is based on the combined signal in the north
and south BBCs.  The charge distribution is fitted using a Monte Carlo
(MC) Glauber~\cite{Loizides:2014vua} simulation to estimate the number
of participating nucleons ($\npart$) and a negative binomial
distribution to describe the BBC signal for fixed $\npart$.  All
quantities in the present manuscript are reported as a function of
centrality and the corresponding $\npart$ values are shown in
Table~\ref{tab:npart}.

\begin{table}[h]
\caption
{$\npart$ values for various centrality categories.}
\begin{ruledtabular} \begin{tabular}{crrc}
\ \ \ \ \ \ \ \ & Centrality & $\mean{\npart}$ \ \ & \ \ \ \ \ \ \ \ \  \\
\hline
&  0\%--5\% & 350.8 $\pm$ 3.1  & \\
&  5\%--10\% & 301.7 $\pm$ 4.7 & \\
& 10\%--20\% & 236.1 $\pm$ 3.8 & \\
& 20\%--30\% & 167.6 $\pm$ 5.5 & \\
& 30\%--40\% & 115.5 $\pm$ 5.8 & \\
& 40\%--50\% &  76.1 $\pm$ 5.5 & \\
& 50\%--60\% &  47.0 $\pm$ 4.7 & \\
& 60\%--70\% &  26.7 $\pm$ 3.6 & \\
& 70\%--80\% &  13.6 $\pm$ 2.4 & \\
& 80\%--93\% &   6.1 $\pm$ 1.3 & \\
\end{tabular} \end{ruledtabular}
\label{tab:npart}
\end{table}

The main detector used in the analysis is the forward silicon vertex
detector (FVTX). The FVTX~\cite{Aidala:2013vna} is a silicon strip
detector comprising two arms, north and south, covering $1<|\eta|<3$. In
Au$+$Au collisions there is a strong correlation between the total signal
in the BBCs and the total number of tracks in the FVTX. To remove beam
related background, we apply an additional event selection on the
correlation between the total BBC signal and the number of tracks in the
FVTX.

\begin{figure}[hbtp]
\includegraphics[width=1.0\linewidth]{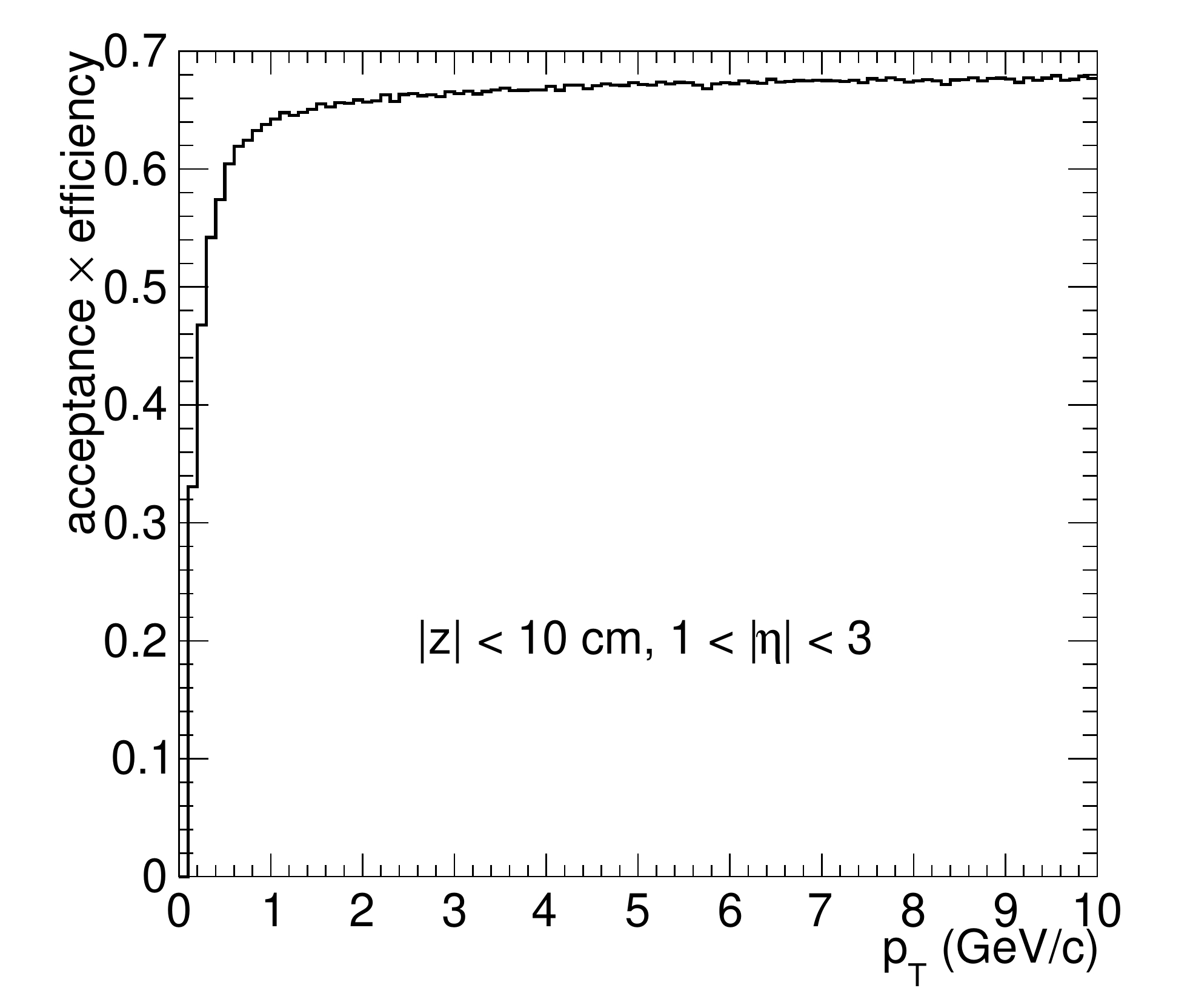}
\caption
{Tracking efficiency and acceptance in the FVTX as a function of $p_T$.
At $p_T$~=~0.3~GeV/$c$ the efficiency is 75\% of its asymptotic value.}
\label{fig:accpt}
\end{figure}

\begin{figure}[hbtp]
\includegraphics[width=1.0\linewidth]{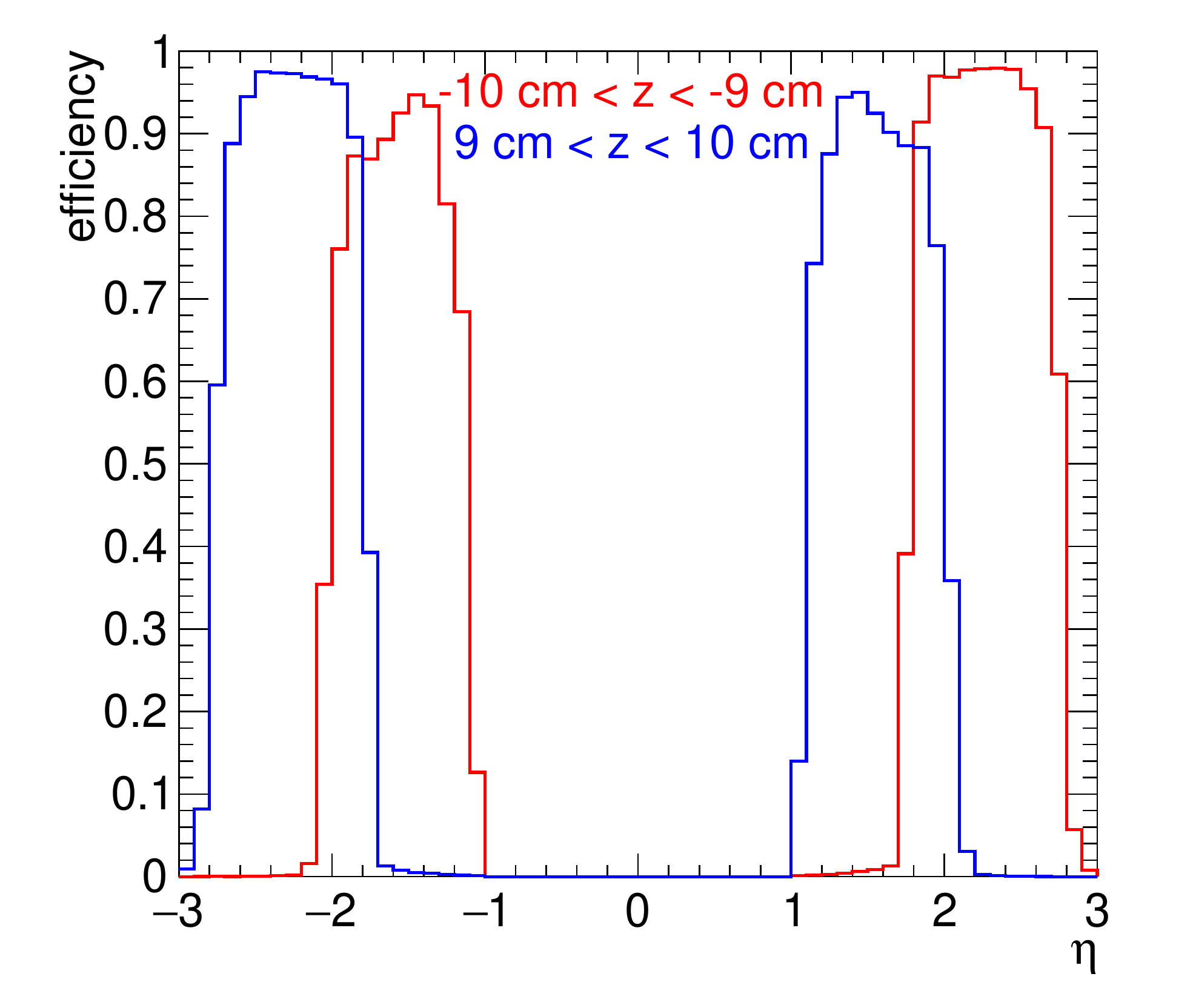}
\caption
{Tracking efficiency in the FVTX as a function of $\eta$ for two
different $z$-vertex selections.}
\label{fig:acceta}
\end{figure}

Each FVTX arm has four layers.  In the track reconstruction software, a
minimum of three hits is required to reconstruct a track.  However, it
is possible for there to be hit sharing with the central rapidity
detector (VTX), so that one or two of the three required hits can be in
the VTX.  We select tracks using a stricter requirement of at least
three hits in FVTX, irrespective of the number of hits in the VTX.  We
further require that the track reconstruction algorithm have a goodness
of fit of $\chisq<5$ for each track. Lastly, we require that each track
has a distance of closest approach (DCA) of less than 2~cm. The DCA is
defined as the distance between the event vertex and the straight-line
extrapolation point of the FVTX track onto a plane which is
perpendicular to the $z$-axis and contains the event vertex. A 2~cm cut
selects the FVTX tracks that likely originate from the event vertex, and
is conservative in accepting the nonzero DCA tail that stems from the
uncertainty in the determination of the vertex position and the bending
of the actual track in the experimental magnetic field. Due to the
orientation of the FVTX strips relative to the magnetic field, momentum
determination is not possible using the tracks in the FVTX alone.
However, \geant~\cite{Agostinelli:2002hh} simulations have determined
that the tracking efficiency is relatively independent of momentum for
$p_T\gtrsim$~0.3~GeV/$c$. Figure~\ref{fig:accpt} shows the $p_T$
dependence of the FVTX tracking efficiency averaged over $1<|\eta|<3$.
Figure~\ref{fig:acceta} shows the tracking efficiency as a function of
$\eta$ in the FVTX for two different $z$-vertex selections. The single
particle tracking efficiency has a maximum value of 98.6\% as a function
of $\eta$. When averaging over $1 < |\eta| < 3$, the maximum value of
the $p_T$-dependent efficiency is 67.9\%, and $p_T$ = 0.3 GeV/$c$ the
efficiency is at 75\% of its maximum value.

\section{Analysis Methods}
\label{sec:analysis}

The azimuthal distribution of particles in an event can be represented
by a Fourier series~\cite{Voloshin:1994mz}:
\begin{equation}
\frac{dN}{d\phi} \propto 1 + \sum_n 2v_n \cos(n(\phi-\psi_n)),
\end{equation}
where $n$ is the harmonic number, $\phi$ is the azimuthal angle of some
particle, $\psi_n$ is the symmetry plane, and $v_n =
\mean{\cos(n(\phi-\psi_n))}$. There are many experimental techniques for
estimating the $v_n$ coefficients, some of which we discuss in this
section.

The main ingredient in the present analysis, for both the cumulant
results and the folding results, is the Q-vector.  The Q-vector is a
complex number $Q_n = Q_{n,x} + iQ_{n,y}$ with the components defined as
\begin{align}
\mathfrak{R} Q_n &= Q_{n,x} = \sum_i^N \cos n\phi_i, \\
\mathfrak{I} Q_n &= Q_{n,y} = \sum_i^N \sin n\phi_i,
\end{align}
where $\phi_i$ is the azimuthal angle of some particle and $N$ is the
number of particles in some event or subevent---a subevent is a subset
of a whole event, usually selected based on some kinematic selection,
e.g. pseudorapidity.  Because the PHENIX FVTX detector subsystem is split
into two separate arms, north ($1<\eta<3$) and south ($-3<\eta<-1$), it
is natural to use tracks in the two arms as separate subevents for some
calculations.  In other calculations, all tracks from north and south
will be combined into a single event.

Additional corrections to the data are needed to account for any
nonuniformity in the azimuthal acceptance. In the case of uniform
azimuthal acceptance, the event average of the Q-vector components is
zero: $\mean{Q_{n,x}} = \mean{Q_{n,y}} = 0$.  In the case of nonuniform
acceptance, there can be a systematic bias such that this relation does
not hold.  In the case of few-particle correlations, i.e. 2- and
4-particle correlations, the bias can be corrected analytically in a
straightforward manner~\cite{Bilandzic:2010jr}. In the case of
correlations with a larger number of particles, however, this becomes
impractical.  The total number of terms in a $k$-particle cumulant
calculation without the assumption of azimuthal uniformity is given by
the Bell sequence: $1,2,5,15,52,203,877,4140,...$---that is, the number
of terms for 2- and 4-particle correlations is a rather manageable 2 and
15, respectively; contrariwise, the number of terms for 6- and
8-particle correlations is a rather unmanageable 203 and 4140 terms,
respectively. For that reason, the only practicable choice is to perform
calculations on corrected Q-vectors. The present analysis makes use of
Q-vector re-centering~\cite{Poskanzer:1998yz}. In this procedure one has
the relation
\begin{equation}
Q^{\rm corrected}_{n} = Q^{\rm raw}_{n} - Q^{\rm average}_{n},
\end{equation}
where
\begin{align}
Q^{\rm average}_{n} = N\mean{\cos n\phi} + iN\mean{\sin n\phi},
\end{align}
and
\begin{align}
\mean{\cos n\phi} &= \mean{Q_{n,x}/N}, \\
\mean{\sin n\phi} &= \mean{Q_{n,y}/N}.
\end{align}

In the present analysis, we perform the Q-vector re-centering procedure
for each FVTX arm separately and as a function of $\nfvtxt$. To assess
the associated systematic uncertainty, we perform the Q-vector
re-centering as a function of centrality instead, as a function of
additional secondary variables (event vertex and operational time
period), and for combined arms instead of separate.

\subsection{Cumulants}
\label{sec:cumulants}

The cumulant method for flow analysis was first proposed in
Ref.~\cite{Borghini:2000sa}.  In the present analysis, we use the
recursion algorithm developed in Ref.~\cite{Bilandzic:2013kga}, which is
a generalization of the direct calculations using Q-vector algebra first
derived in Ref.~\cite{Bilandzic:2010jr}. We consider 2-, 4-, 6-, and 8-
particle correlations. The multi-particle correlations are denoted
$\mean{k}$ for $k$-particle correlations and are estimators for the
$k$-th moment of $v_n$, i.e. $\mean{k} = \mean{v_n^k}$. In terms of the
angular relationships between different particles, they are
\begin{align}
\mean{2} &= \mean{\cos(n(\phi_1-\phi_2))}, \\
\mean{4} &= \mean{\cos(n(\phi_1+\phi_2-\phi_3-\phi_4))}, \\
\mean{6} &= \mean{\cos(n(\phi_1+\phi_2+\phi_3-\phi_4-\phi_5-\phi_6))}, \\
\mean{8} &= \mean{\cos(n(\phi_1+\phi_2+\phi_3+\phi_4-\phi_5-\phi_6-\phi_7-\phi_8))},
\end{align}
where $\phi_{1,...,8}$ represent the azimuthal angles of different
particles in the event.

The $k$-particle cumulants, denoted $c_n\{k\}$ are constructed in such a
way that potential contributions from lower order correlations are
removed.  Because the cumulants mix various terms that are of equal powers
of $v_n$, the cumulant method $v_n$, denoted $v_n\{k\}$, is proportional
to the $k$-th root of the cumulant. The $c_n\{k\}$ are constructed as
follows:
\begin{align}
\cnt &= \mean{2}, \\
\cnf &= \mean{4} - 2\mean{2}^2, \\
\cnsix &= \mean{6} - 9\mean{4}\mean{2} + 12\mean{2}^3,  \\
\cne &= \mean{8} - 16\mean{6}\mean{2} - 18\mean{4}^2 + 144\mean{4}\mean{2}^2 - 144\mean{2}^4,
\end{align}
and the $v_n\{k\}$ are
\begin{align}
\vnt &= (\cnt)^{1/2}, \\
\vnf &= (-\cnf)^{1/4}, \\
\vns &= (\cnsix/4)^{1/6}, \\
\vne &= (-\cne/33)^{1/8}.
\end{align}

It is also possible to construct cumulants in two or more subevents,
though in the present analysis we will only concern ourselves with two
subevents.  For 2-particle correlations, rather than $\mean{2} =
\mean{\cos(n(\phi_1-\phi_2))}$ where $\phi_1$ and $\phi_2$ are from the
same subevent, one has instead $\mean{2}_{a|b} =
\mean{\cos(n(\phi_1^a-\phi_2^b))}$ where $a$, $b$ denote two different
subevents.  The cumulant and $v_n$ have the same relationship as in the
single event case, i.e. $\vnt_{a|b}^2 = \cnt_{a|b} = \mean{2}_{a|b}$.
The two-subevent 2-particle cumulant is also known as the scalar product
method~\cite{Adler:2002pu}.

Subevent cumulants for correlations with four or more particles were
first proposed in Ref.~\cite{Jia:2017hbm}.  For two-subevent 4-particle
correlations, there are two possibilities:
\begin{align}
\mean{4}_{ab|ab} &= \mean{\cos(n(\phi_1^a+\phi_2^b-\phi_3^a-\phi_4^b))}, \\
\mean{4}_{aa|bb} &= \mean{\cos(n(\phi_1^a+\phi_2^a-\phi_3^b-\phi_4^b))},
\end{align}

where the former allows 2-particle correlations within single subevents
and the latter excludes them.  The latter is therefore less susceptible
to nonflow than the former, although both are less susceptible to
nonflow than single event 4-particle correlations. The cumulants take
the form
\begin{align}
\cnf_{ab|ab} &= \mean{4}_{ab|ab} - \mean{2}_{a|a}\mean{2}_{b|b} - \mean{2}_{a|b}^2, \\
\cnf_{aa|bb} &= \mean{4}_{aa|bb} - 2\mean{2}_{a|b}^2,
\end{align}
and the $\vnf$ values have the same relationship to the cumulants as in
the single particle case, i.e. $\vnf_{ab|ab} = (-\cnf_{ab|ab})^{1/4}$
and $\vnf_{aa|bb} = (-\cnf_{aa|bb})^{1/4}$.

To determine systematic uncertainties associated with event and track 
selection for the cumulant analysis, we vary the event and track 
selection criteria and assess the variation on the final analysis 
results. The $z$-vertex selection is modified from $\pm$~10~cm to 
$\pm$~5~cm. The track selections are independently modified to have a 
goodness of fit requirement $\chisq<3$. These changes move the 
cumulant results by an almost common multiplicative value and thus we 
quote the systematic uncertainties as a global scale factor 
uncertainty for each result.

\subsection{Folding}
\label{sec:folding}

Here we describe an alternative approach where one utilizes the 
event-by-event $Q_{n}$ distribution to extract the event-by-event 
$v_{n}$ distribution via an unfolding procedure. For our analysis we 
attempt a procedure similar to that used by ATLAS as described in 
Ref.~\cite{Aad:2013xma}.

In brief, ATLAS successfully carries out the unfold in Pb$+$Pb collisions
at 2.76~TeV and finds that the event-by-event probability distribution
for elliptic flow $p(v_{2})$ is reasonably described by a
Bessel-Gaussian function
\begin{equation}
p(v_n) = \frac{v_n}{\delta_{v_n}^2} e^{-\frac{(v_n^2 + (v_n^{\rm RP})^2)}{2\delta_{v_n}^2}} I_0\left(\frac{v_nv_n^{\rm RP}}{\delta_{v_n}^2}\right),
\label{eqn:prob}
\end{equation}
where $v_n^{\rm RP}$ and $\delta_{v_n}^2$ are function parameters that
are related \textit{but not equal} to the mean and variance of the
distribution, respectively. Because flow is a vector quantity, it has
both a magnitude and a phase.  When measuring $v_n$ one measures the
modulus of the complex number, meaning there is a reduction in the
number of dimensions from two to one.  If the fluctuations in each
dimension are Gaussian, one then expects the final distribution of
values to be Bessel-Gaussian.  Recently the CMS experiment has carried
out a similar flow unfolding and observes small deviations from the
Bessel-Gaussian form, favoring the elliptic power
distribution~\cite{Sirunyan:2017fts}.

For the unfolding, ATLAS determines the response matrix in a data driven
way.  The smearing in the response matrix is modest as Pb$+$Pb collisions
have a high multiplicity and the ATLAS detector has large phase space
coverage for tracks $-2.5 < \eta < +2.5$. In our case, the multiplicity
of Au$+$Au collisions is lower in comparison with the multiplicity in
Pb$+$Pb collisions and the phase space coverage of the FVTX detector is
significantly smaller.  Hence, the smearing as encoded in the response
matrix is significantly larger and the unfolding is more challenging.

\begin{figure*}[hbtp]
\includegraphics[width=0.99\linewidth]{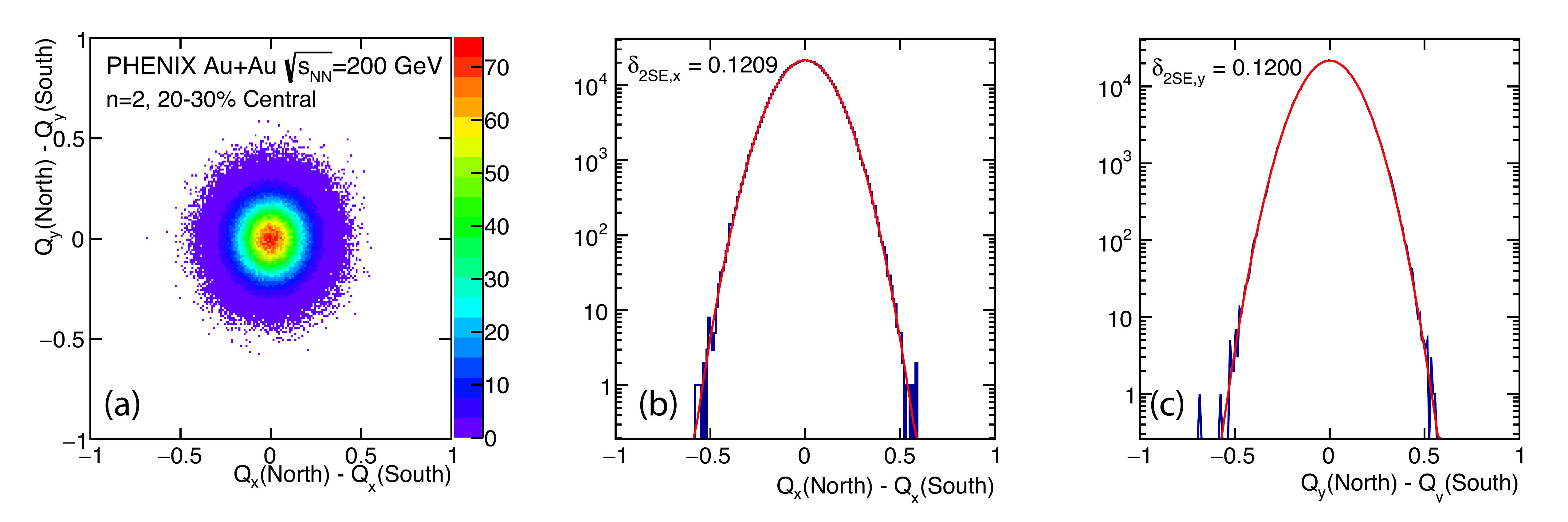}
\caption
{Example distribution of $Q^{\rm north} - Q^{\rm south}$ for the $n=2$
case corresponding to \auau collisions at $\snn$ = 200 GeV and
centrality 20--30\%. (a) The two-dimensional distribution.  (b) The
projection onto $Q_x$.  (c) The projection onto $Q_y$.  Shown for (b)
and (c) are the extracted Gaussian widths $\delta_{2SE}$---the $\chisq$
values of the fits are 1.02 and 1.18 for (b) and (c), respectively.
}
\label{fig:qdiffexample}
\end{figure*}

To estimate the response function of the detector, we follow the
procedure from ATLAS~\cite{Aad:2013xma}, which is to examine the
difference between two subevents for both $Q_x$ and $Q_y$. We compare
the $Q$-vector determined in the south arm of the FVTX, $Q^{\rm south}$,
to the $Q$-vector determined in the north arm of the FVTX, $Q^{\rm
north}$. Figure~\ref{fig:qdiffexample} shows an example of this
procedure for the $n=2$ case. Figure~\ref{fig:qdiffexample}~(a) shows
the 2-dimensional distribution for the 20\%--30\% centrality selection;
Fig~\ref{fig:qdiffexample}~(b) shows the one-dimensional projection of
this onto the $x$-axis (i.e. the one-dimensional distribution of
$Q_x^{\rm north} - Q_x^{\rm south}$); Fig~\ref{fig:qdiffexample}~(c)
shows the one-dimensional projection of this onto the $y$-axis (i.e. the
one-dimensional distribution of $Q_y^{\rm north} - Q_y^{\rm south}$).
These distributions in all centrality selections are Gaussian over four
orders of magnitude, and we characterize them via their Gaussian widths
$\delta_{2SE}$ which are given in Table~\ref{tab:deltareso}.  It is
notable that these widths are more than a factor of two larger than
those quoted by ATLAS in Pb$+$Pb collisions, for example $\delta_{2SE} =
0.050$ for Pb$+$Pb 20\%--25\% central events~\cite{Aad:2013xma}.

\begin{table}[h]
\caption
{Resolution parameter $\delta_{2SE}$ values for $n=2$ and $n=3$
for various centrality categories.}
\begin{ruledtabular} \begin{tabular}{crccc}
\ \ \ \ \ \ & Centrality & $\delta_{2SE} (n=2)$ & $\delta_{2SE} (n=3)$ & \ \ \ \ \ \  \\
\hline
&   0\%--5\% & 0.117 & 0.115 & \\
&  5\%--10\% & 0.115 & 0.113 & \\
& 10\%--20\% & 0.115 & 0.113 & \\
& 20\%--30\% & 0.121 & 0.119 & \\
& 30\%--40\% & 0.133 & 0.130 & \\
& 40\%--50\% & 0.154 & 0.151 & \\
\end{tabular} \end{ruledtabular}
\label{tab:deltareso}
\end{table}

If there is a modest longitudinal decorrelation between the two
subevents, it will manifest as a slight increase in the $\delta_{2SE}$
parameter.  The final $v_n$ is averaged over that decorrelation.  This
effect, as in previous unfolding analyses~\cite{Aad:2013xma}, is neglected.

We highlight that even in the case of a perfect detector with perfect
acceptance, there remains a smearing due to the finite particle number
in each event.  This raises a question regarding the meaning of a true
$v_{n}$ that is being unfolded back to.  In a hydrodynamic description,
there is a continuous fluid from which one can calculate a single true
anisotropy $v_n$ for each event.  If one then has the fluid breakup into
a finite number of particles $N$, e.g. via Cooper-Frye freeze-out, the
anisotropy of those $N$ particles will fluctuate around the true fluid
value.  However, in a parton scattering description, for example
\ampt~\cite{Lin:2004en}, the time evolution is described in terms of a
finite number of particles $N$.  In this sense there is no separating of
a true $v_n$ from that encoded in the $N$ particles themselves.
Regardless, one can still mathematically apply the unfolding and compare
experiment and theory as manipulated through the same algorithm.

As noted before, the one-dimensional radial projection of a
two-dimensional Gaussian is the so-called Bessel-Gaussian function.  In
this case it means that the conditional probability to measure a value
$v_n^{\rm obs}$ given a true value $v_n$ has the following
Bessel-Gaussian form:
\begin{equation}
p(v_n^{\rm obs}|v_n) \propto v_n^{\rm obs} e^{-\frac{(v_n^{\rm obs})^2 + v_n^2}{2\delta^2}} I_0\left(\frac{v_n^{\rm obs}v_n}{\delta^2}\right),
\label{eqn:response}
\end{equation}
where $\delta$ is the smearing parameter characterizing the response due
to the finite particle number (including from the detector efficiency
and acceptance), and $I_0$ is a modified Bessel function of the first
kind.  The smearing parameter $\delta$ uses the combination of the two
FVTX arms and is related to the result from the difference by $\delta =
\delta_{2SE}/2$.  We highlight that the Bessel-Gaussian in
Eqn.~\ref{eqn:response} is different from the Bessel-Gaussian in
Eqn.~\ref{eqn:prob}, though both arise from a similar dimensional
reduction.

\begin{figure*}[hbtp]
\includegraphics[width=0.98\linewidth]{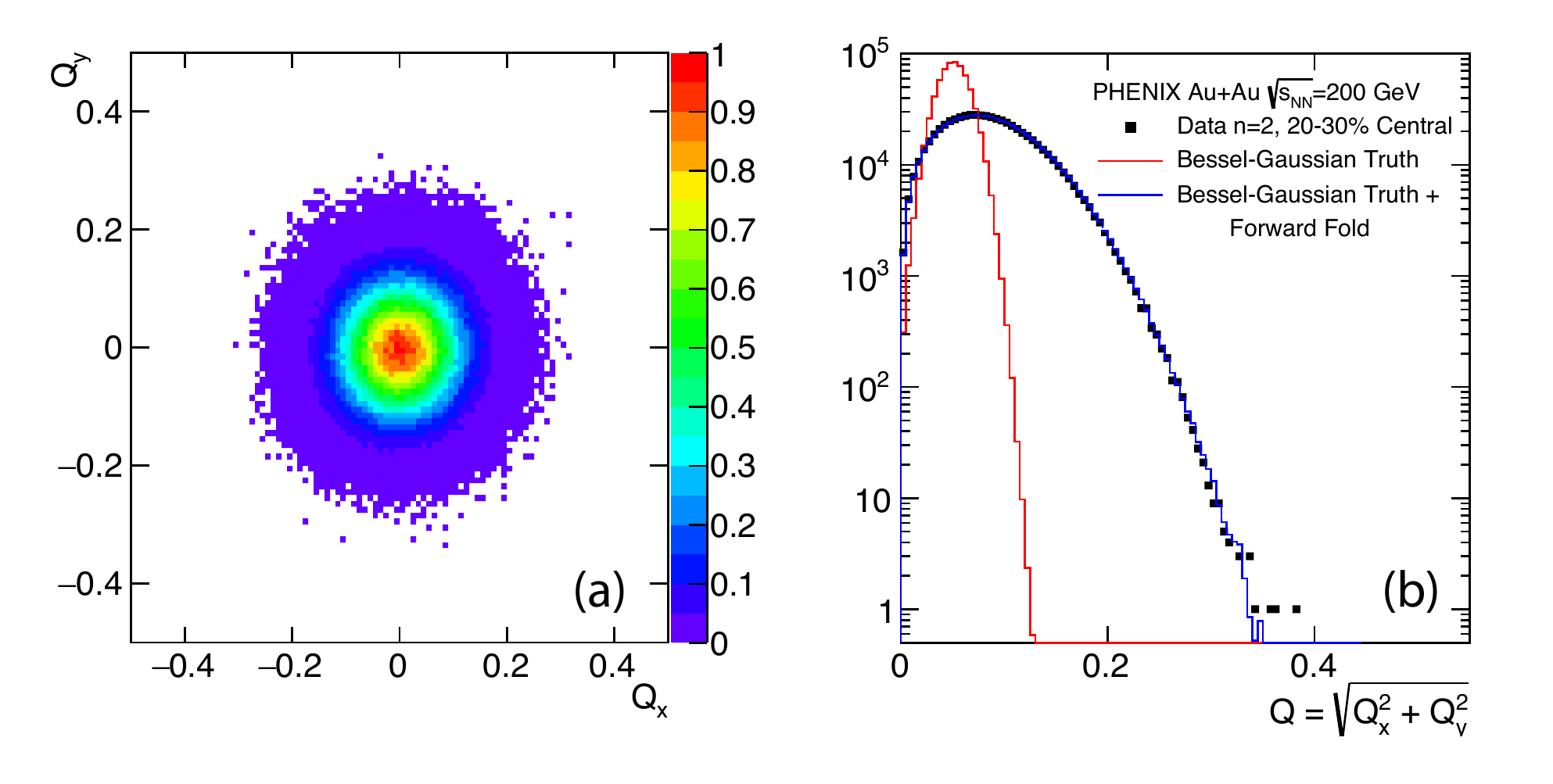}
\includegraphics[width=0.98\linewidth]{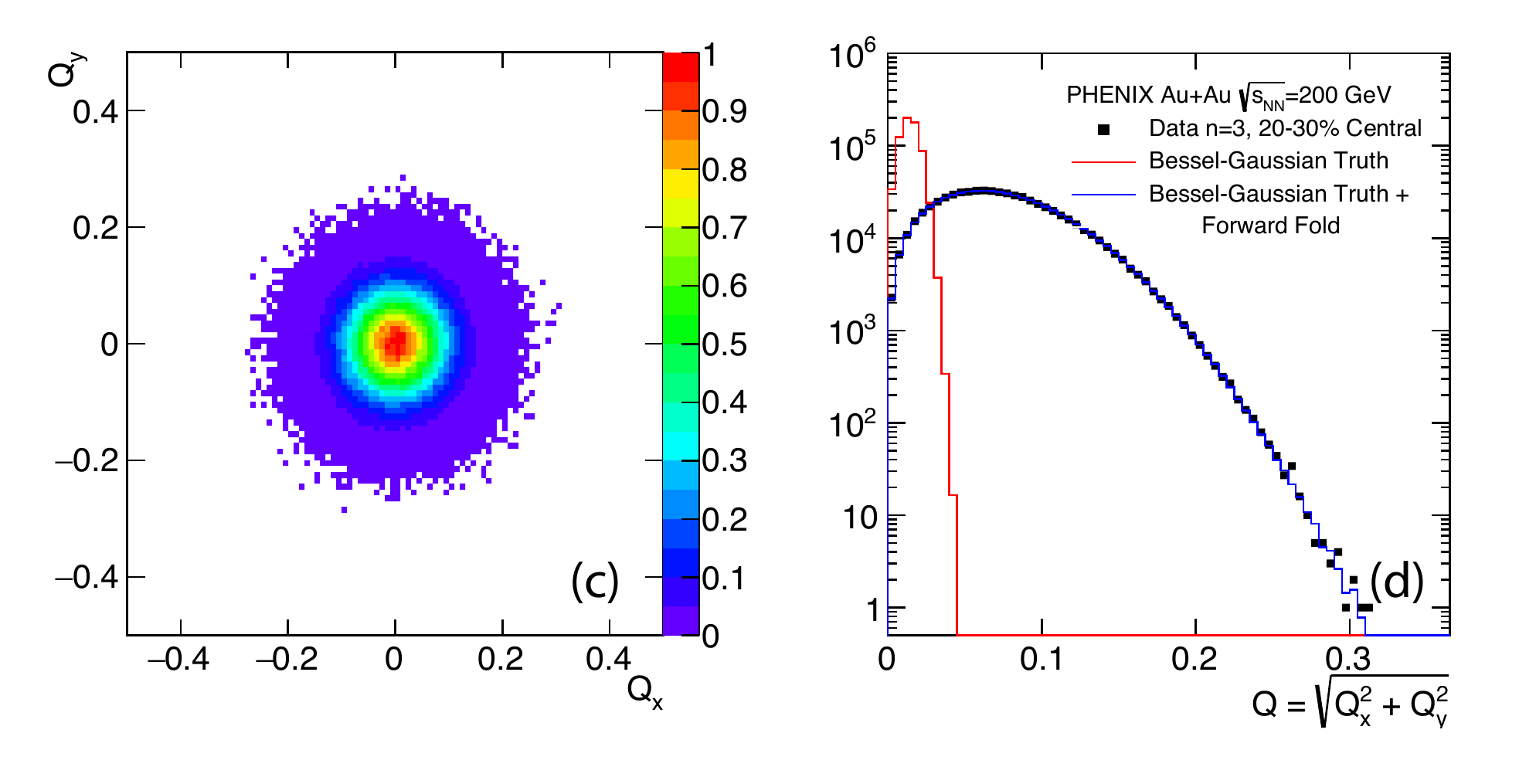}
\caption
{Example distribution of $Q$ for the (a), (b) elliptic $n=2$ case and
(c), (d) triangular $n=3$ case (lower). (a) and (c) show the
2-dimensional distribution. (b) and (d) show the 1-dimensional
distribution $|Q|$. The distribution corresponds to \auau collisions at
$\snn$ = 200 GeV and centrality 20--30\%. Also shown in (b), (d) is the
best fit for this run Bessel-Gaussian truth distribution and the
corresponding forward folded result (i.e. pushing the truth distribution
through the response matrix).
}
\label{fig:qtotal}
\end{figure*}

We have employed the iterative Bayes unfold method as encoded in
RooUnfold~\cite{Adye:2011gm} and our own implementation of a Singular
Value Decomposition (SVD) unfold
method~\cite{Hansen:1990,Hocker:1995kb}.  In both cases, the response
matrix is populated using Eqn.~\ref{eqn:response} using the
data-determined smearing parameters.  The FVTX-determined event-by-event
$Q_n$ distributions are used as input to the unfold.
Figure~\ref{fig:qtotal} shows this dimensional reduction for the $n=2$
and $n=3$ case, respectively. Figure~\ref{fig:qtotal} (a) and (c) show
the two-dimensional distribution of $Q_n$ and Figure~\ref{fig:qtotal}
(b) and (d) show the one-dimensional distribution of $|Q_{n}|$ for the
20\%--30\% centrality range.

The Au$+$Au 20\%--30\% centrality class is expected to provide the best
conditions, in terms of the predicted $\mean{v_2}$ and resolution
$\delta$, to determine $p(v_{2})$ via unfolding. However, because it is
quite challenging to unfold the measured distribution directly, we
constructed a test version of the problem to illustrate the procedure
using the SVD method.  The details of this test are given in
Appendix~\ref{APPENDIX}, but the end result is that the unfolding
procedure is inherently unstable and therefore fails to converge for the
resolution parameters in the present analysis.

As a result, instead of inverting the response matrix, we can make an
{\it{ansatz}} that the probability distributions, $p(v_{2})$ and
$p(v_{3})$ are exactly Bessel-Gaussian in form.  Under this restrictive
assumption, because the Bessel-Gaussian form has only two parameters as
shown in Eqn.~\ref{eqn:prob}, we can simply evaluate a large grid of
parameter combinations as guesses for the truth distribution and forward
fold them, i.e. passing them through the response matrix to compare to
the observed $Q_n$ distribution.  We have carried out such a ``forward
fold'' procedure with over 10,000 parameter combinations.  We then
determine the statistical best fit parameters and their statistical
uncertainties based on a $\chi^{2}$ mapping. We consider only the
Gaussian statistical uncertainties here, and detail our treatment of
systematic uncertainties in Section~\ref{sec:foldingresults}. Some
advantages of this procedure are that we explore the full $\chi^2$ space
and have no sensitivity to an unfolding prior, regularization scheme,
and number of iterations.  The disadvantage of course is the
{\it{ansatz}} that the distribution is precisely Bessel-Gaussian.

\begin{figure*}[hbtp]
\includegraphics[width=0.98\linewidth]{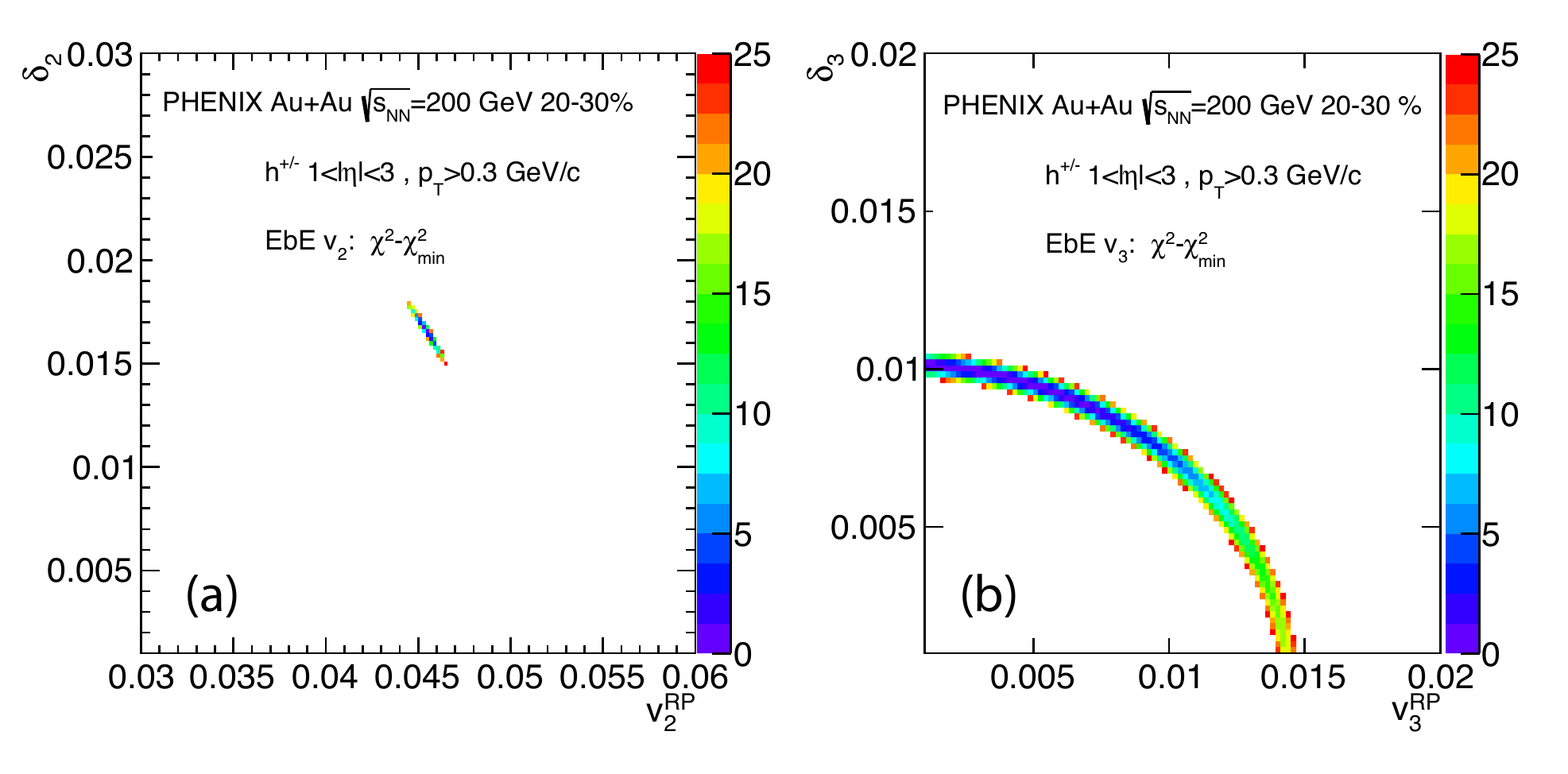}
\caption
{Two dimensional color plot of $\chi^2 - \chi_{\rm min}^{2}$ values as a
function of the two Bessel-Gaussian parameters, $v_n^{\rm RP}$ and
$\delta_{v_n}$ for the 20\%--30\% centrality bin. (a) shows the second
harmonic and (b) shows the third harmonic.  Only $\chi^{2}$ values up to
$\chi_{\rm min}^{2} + 25$ are shown.
}
\label{fig:chi2}
\end{figure*}

Examples of this $\chi^{2}$ forward fold mapping are shown in
Figure~\ref{fig:chi2}.  It is striking that for the $v_2$ case in the
Au$+$Au 20\%--30\% central bin, the forward folding reveals a tight
constraint on the Bessel-Gaussian parameters.  In contrast, for the same
centrality bin and $v_3$, there is a band of parameter combinations
providing a roughly equally good match to the experimental $Q_3$
distribution.  Shown in Figure~\ref{fig:qtotal} for $v_2$ (upper) and
$v_{3}$ (lower) are the best Bessel-Gaussian fit distributions (red) and
their forward folded results (blue).  Both cases show good agreement
between the forward folded results and the measured experimental
distribution.  The corresponding best $\chi^{2}_{\rm min}$ values indicate a good
match.  It is notable that in the $v_{3}$ case, the $\chi^{2}_{\rm min}$ values
are slightly worse in all cases and significantly worse in the most
peripheral selection.  The more peripheral data have a slightly larger tail at high
$Q_3$ values which could indicate an incompatibility with the
Bessel-Gaussian {\it{ansatz}}.

\section{Results and Discussion}
\label{sec:discussion}

Here we detail the full set of results for the elliptic and triangular
flow moments and distributions in Au$+$Au collisions at $\snn$ = 200 GeV.
We start by detailing the cumulant results.

\subsection{Cumulants Results}

\begin{figure*}[hbtp]
\includegraphics[width=0.49\linewidth]{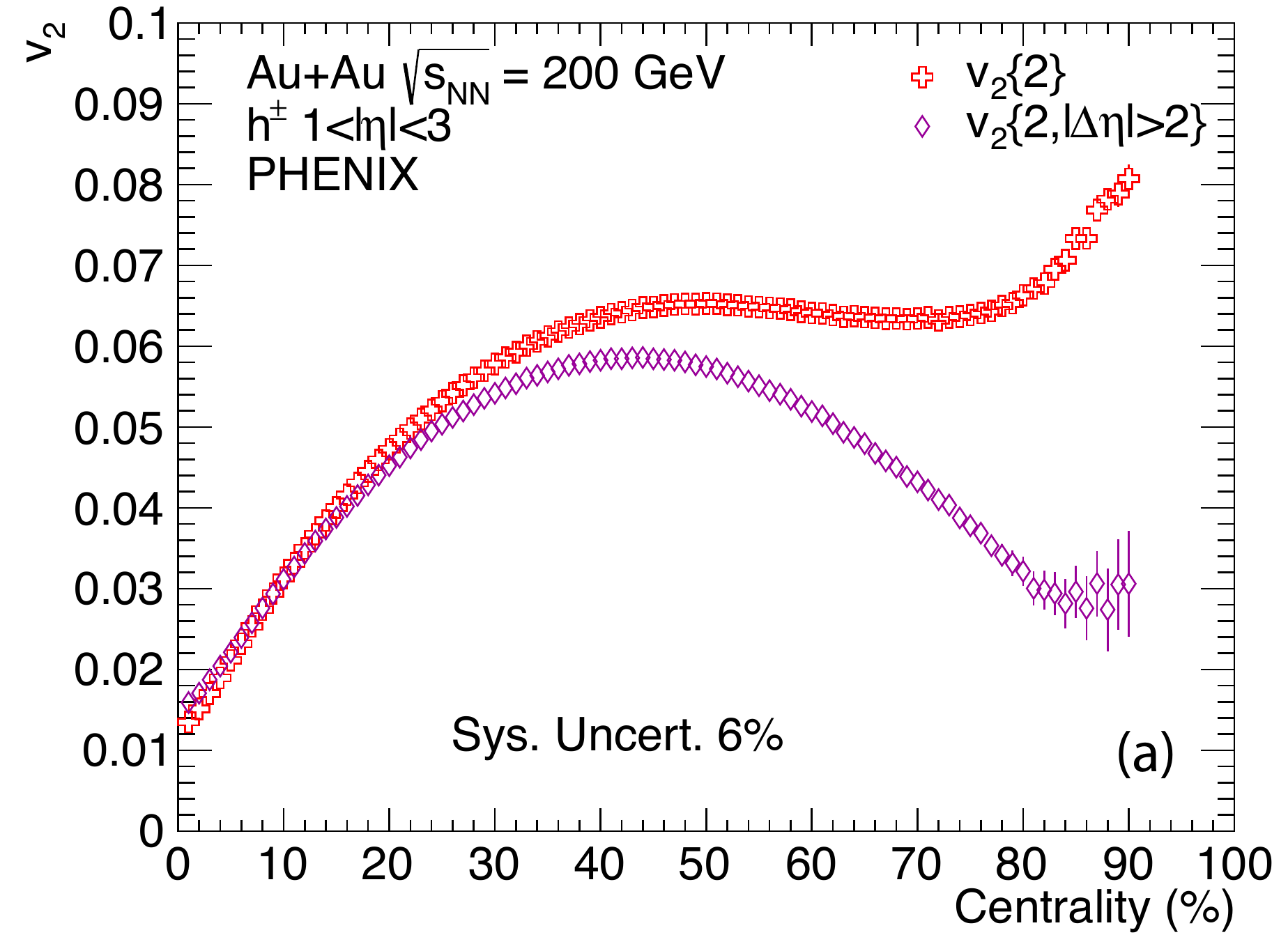}
\includegraphics[width=0.49\linewidth]{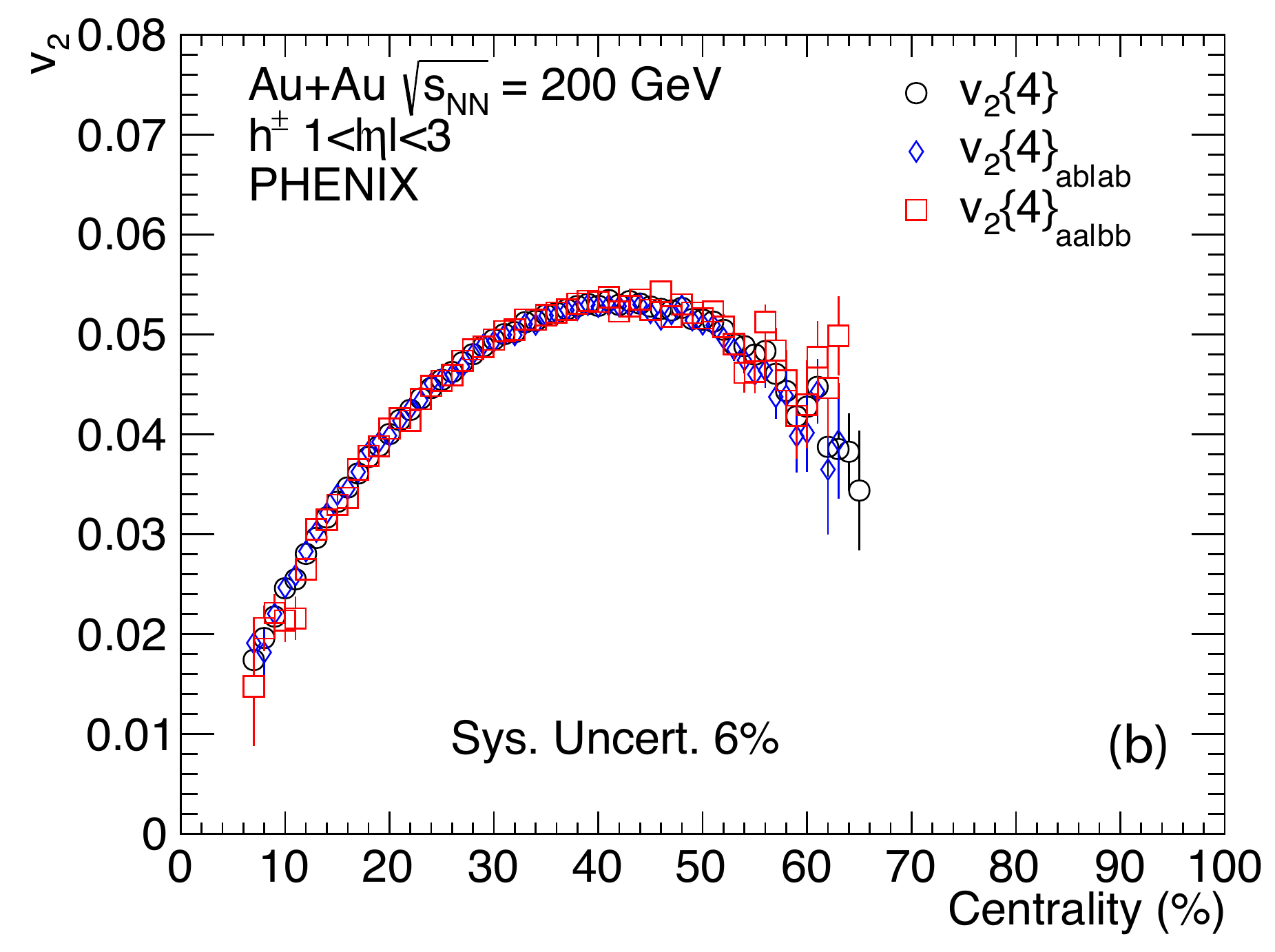}
\caption
{Centrality dependence of (a) $\vtt$ and (b) $\vtf$. (a) The red points
indicate no pseudorapidity gap whereas the magenta points indicate a
pseudorapidity gap of $|\Delta\eta|>$~2.0. (b) The black points indicate
$\vtf$ with no pseudorapidity gap, the blue points indicate a
two-subevent method with $|\Delta\eta|>$~2.0 but where some short-range
pairs are allowed, and the red points indicate a two-subevent method
with $|\Delta\eta|>$~2.0 where no short-range pairs are allowed.
}
\label{fig:v2sub}
\end{figure*}

First we show in Figure~\ref{fig:v2sub} (a) and (b) the centrality
dependence of $\vtt$ and $\vtf$, respectively.  The statistical
uncertainties are shown as vertical lines and the systematic uncertainty
is quoted as a global factor uncertainty. (a) shows a dramatic
difference for centrality larger than 40\% between the red points,
obtained without requiring a pseudorapidity gap in the particle pair,
and the magenta points, which have a pseudorapidity gap of
$|\Delta\eta|>$~2.0. This is due to the fact that the pseudorapidity gap
removes a large amount of nonflow, especially in the peripheral
collisions where nonflow is combinatorially less suppressed relative to
central collisions. Contrariwise, (b) shows no difference between the
black points (no pseudorapidity gap) and two different 2-subevent
cumulants, one where short-range pairs are allowed (blue points) and one
where they are not (red points).  The absence of any effect here
indicates that the 4-particle correlation sufficiently suppresses
nonflow combinatorially such that the kinematic separation of particles
provides no additional benefit.  Note that this is not necessarily the
case in smaller collision systems---subevent cumulants have been shown
to significantly reduce nonflow in $p$+$p/$Pb collisions at the
LHC~\cite{Aaboud:2017blb}, and are of potential
interest in $p/d/^3$He+Au collisions at RHIC.

\begin{figure}[hbtp]
\includegraphics[width=1.0\linewidth]{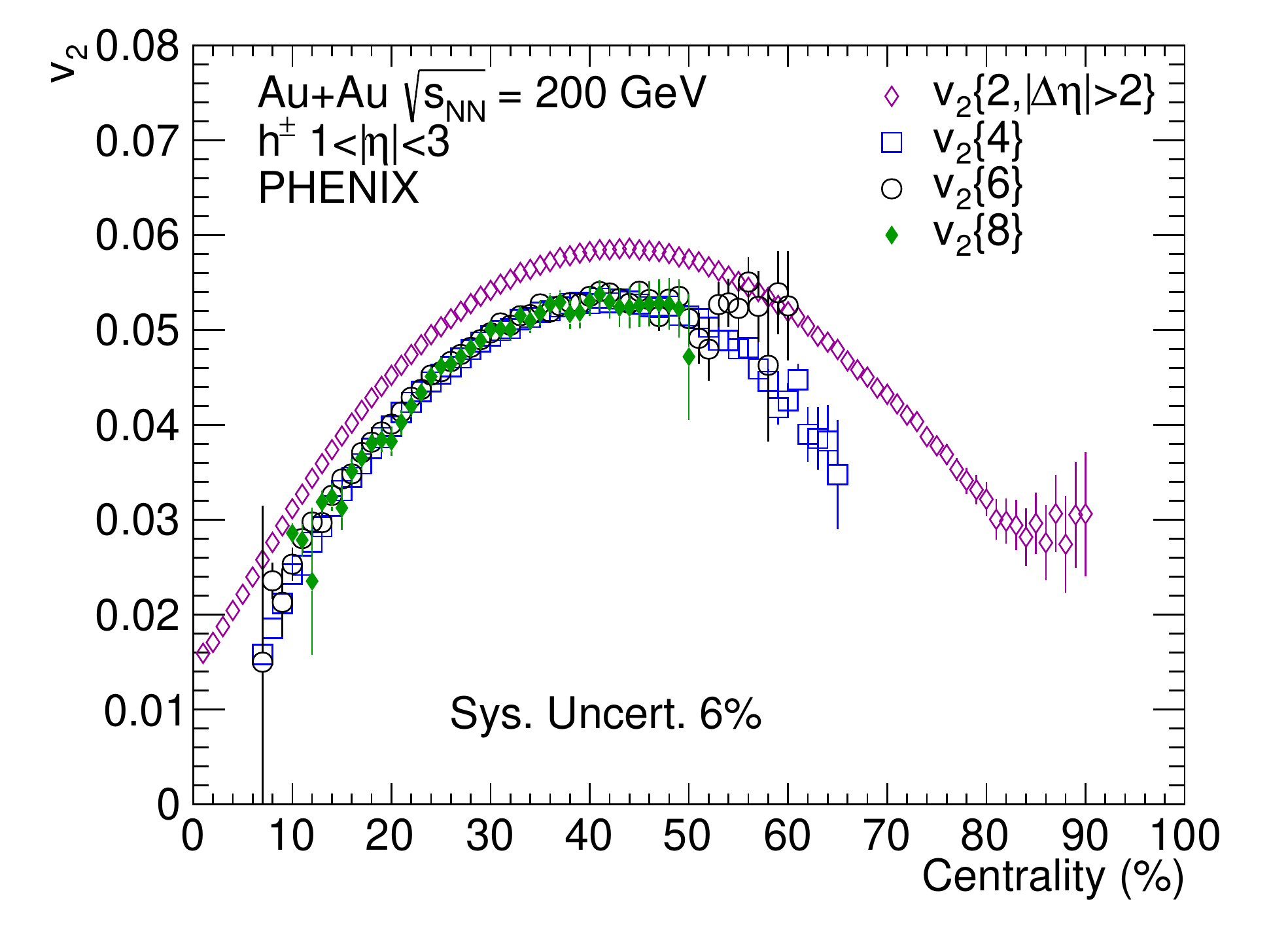}
\caption
{Multi-particle $v_2$ as a function of centrality in Au$+$Au collisions at
$\snn$ = 200 GeV.  The magenta open diamonds indicate the $v_2\{SP\}$,
the blue open squares indicate $\vtf$, the black open circles indicate
$\vts$, and the green filled diamonds indicate $\vte$.
}
\label{fig:v2multi}
\end{figure}

Figure~\ref{fig:v2multi} shows the centrality dependence of
multi-particle $v_2$, with 2, 4, 6, and 8 particles.  The 4-, 6-, and 8-
particles $v_2$ values are consistent with each other, as expected from
the small-variance limit~\cite{Voloshin:2007pc}. When accounting for the
$\eta$-dependence of $v_2$ as measured by PHOBOS~\cite{Back:2004mh},
which indicates that $v_2$ at $1<|\eta|<3$ is about 1.25 times lower
than it is at $|\eta|<1$, the 2-, 4-, and 6-particle cumulant $v_2$ are
in good agreement with the STAR results~\cite{Adams:2004bi}.

\begin{figure}[hbtp]
\includegraphics[width=1.0\linewidth]{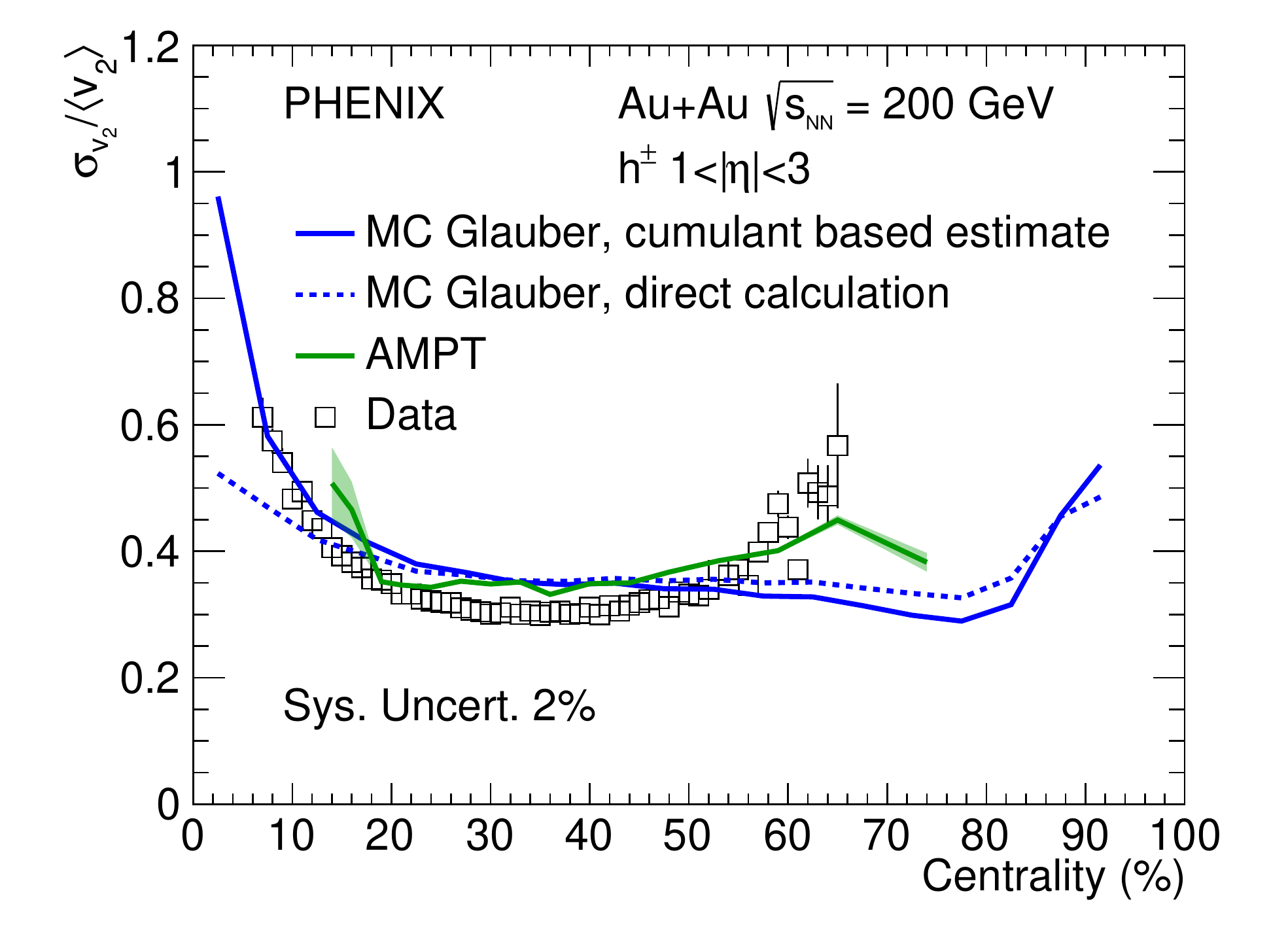}
\caption
{Cumulant method estimate of $\sigmaot$ as a function of centrality in
Au$+$Au collisions at $\snn$ = 200 GeV.  The data are shown as black open
squares.  The same calculation as done in data is done in \ampt, shown
as a solid green line.  Calculations of $\sigma_{\eps_2}/\mean{\eps_2}$
performed in the Monte Carlo Glauber model are shown as blue lines.  The
solid blue line is the Monte Carlo Glauber calculation done using the
same estimate as the data, the dashed blue line is the direct
calculation of the moments of the MC Glauber $\eps_2$ distribution.
}
\label{fig:cumufluc}
\end{figure}

Considering that $\vnt = \sqrt{v_n^2+\sigman^2}$ and that in the small
variance limit
$\vnf \approx \sqrt{v_n^2-\sigman^2}$~\cite{Ollitrault:2009ie}, one can
estimate the relative fluctuations as
\begin{equation}
\frac{\sigman}{\mean{v_n}} \approx \sqrt{ \frac{(\vnt)^2 - (\vnf)^2}{(\vnt)^2 + (\vnf)^2} }.
\end{equation}
Figure~\ref{fig:cumufluc} shows the centrality dependence of this
cumulant estimate of $\sigmaot$. Despite the difference in the rapidity
region where the data are measured, they are in good agreement with
STAR~\cite{Adams:2004bi} and PHOBOS~\cite{Alver:2010rt}. Also shown is a
comparison with \ampt analyzed via cumulants in the same way as the
experimental data.  There is good agreement between the two, indicating
that the Monte Carlo Glauber initial conditions in \ampt and their
fluctuations capture the key event-by-event varying ingredients. We can
also calculate the event-by-event variations in the initial conditions
directly via Monte Carlo Glauber.  In this case we utilize the
event-by-event spatial eccentricity $\varepsilon_{n}$ distributions.  If
there is a linear mapping between initial spatial eccentricity and final
momentum anisotropy ($\varepsilon_{n} \propto v_{n}$), we should expect
a good match between
$\sigma_{\varepsilon_{n}}/\left<\varepsilon_{n}\right>$ and
$\sigma_{v_{n}}/\left<v_{n}\right>$. Also show in
Fig.~\ref{fig:cumufluc} is the Monte Carlo Glauber result via the
calculation of cumulants (solid blue line), as well as the direct
calculation of the variance and mean from the full $\varepsilon_n$
distribution (dashed blue line).  One sees that in midcentral
10\%--50\% collisions, the data and both theory curves agree reasonably.
For more central collisions, the Monte Carlo Glauber data-style
calculation shows the same trend as the data whereas the Monte Carlo
Glauber direct calculation is significantly lower.  This is due to the
fact that the small-variance limit is not a valid approximation in
central collisions.  In peripheral collisions, both Monte Carlo Glauber
curves under-predict the data.  This has been attributed to the
nonlinear response in hydrodynamics~\cite{Noronha-Hostler:2015dbi}.

\begin{figure}[hbtp]
\includegraphics[width=1.0\linewidth]{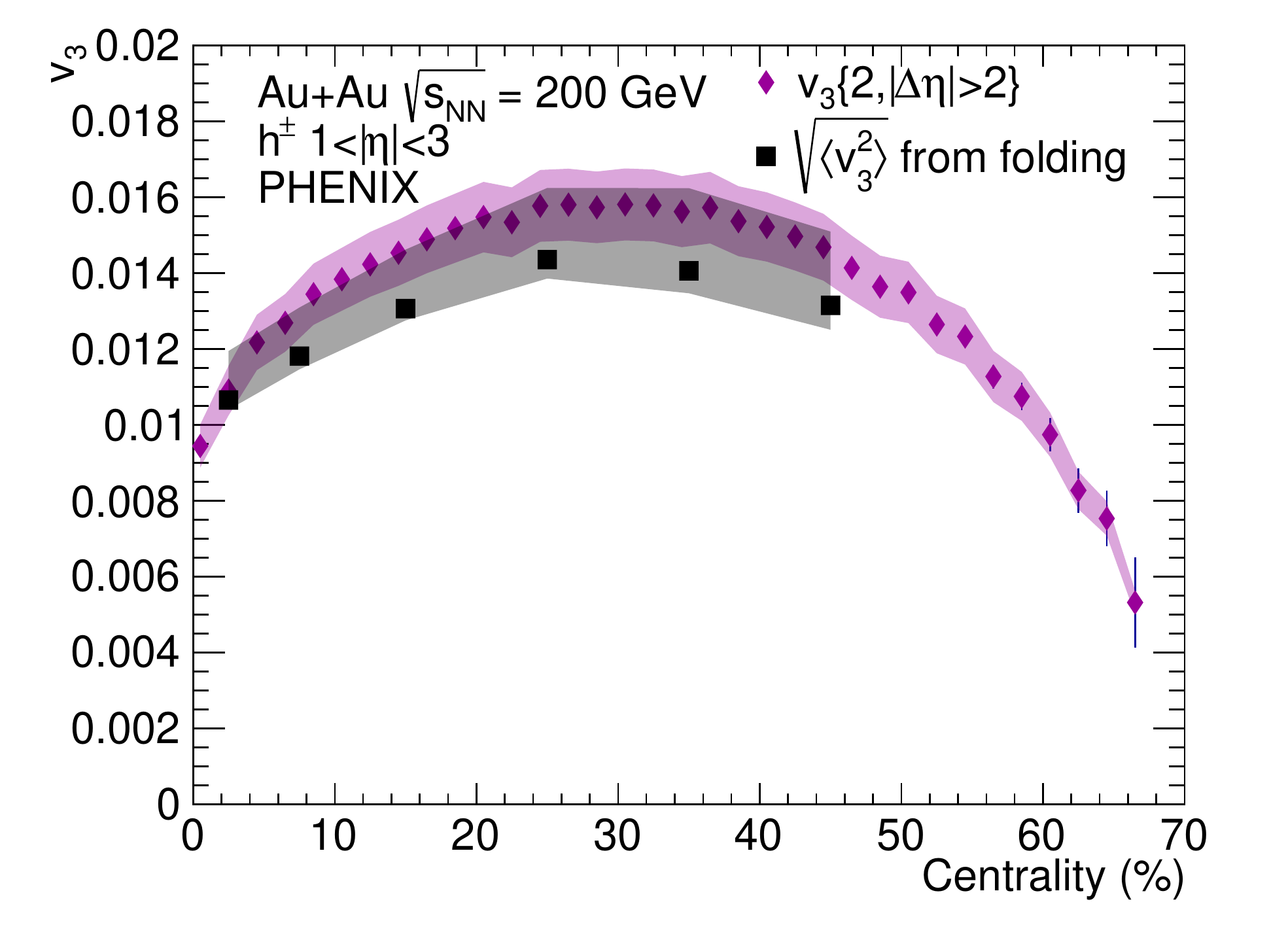}
\caption
{Centrality dependence of $v_3\{2,|\Delta\eta|>2\}$ in Au$+$Au collisions at $\snn$ = 200 GeV,
shown as magenta diamonds.  The systematic uncertainty is indicated as a shaded magenta band.
Also shown as black squares are $\sqrt{\mean{v_3^2}}$ as determined from the folding analysis,
which is shown in the next section.
}
\label{fig:v32s}
\end{figure}

Now we consider the $v_3$ case.  Figure~\ref{fig:v32s} shows
$v_{3}\{2,|\Delta\eta|>2\}$ as a function of centrality.  The centrality
dependence of $v_3$ is much smaller than that of $v_2$, which is
expected because triangular flow is generated dominantly through
fluctuations.

\begin{figure*}[hbtp]
\includegraphics[width=0.49\linewidth]{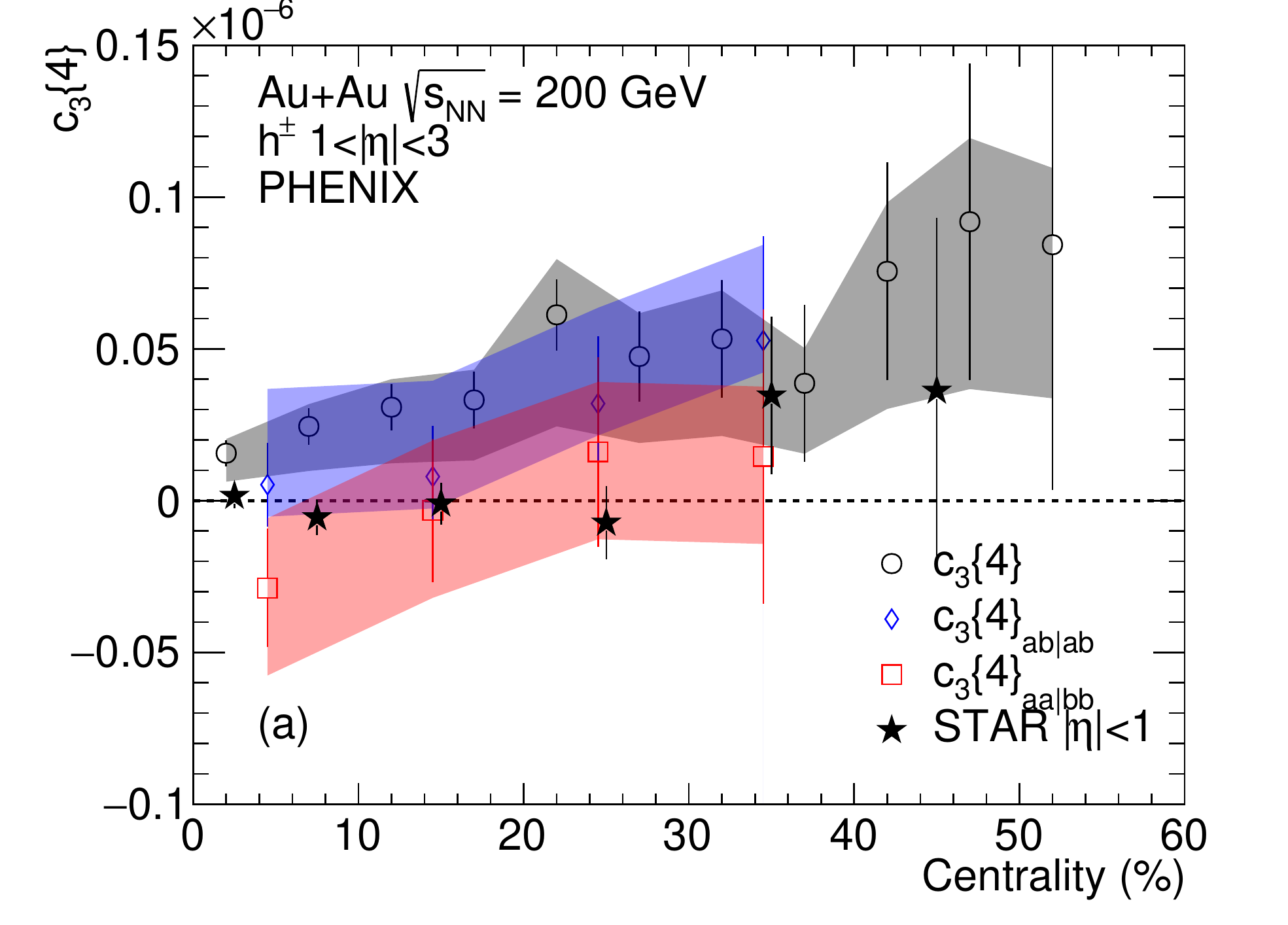}
\includegraphics[width=0.49\linewidth]{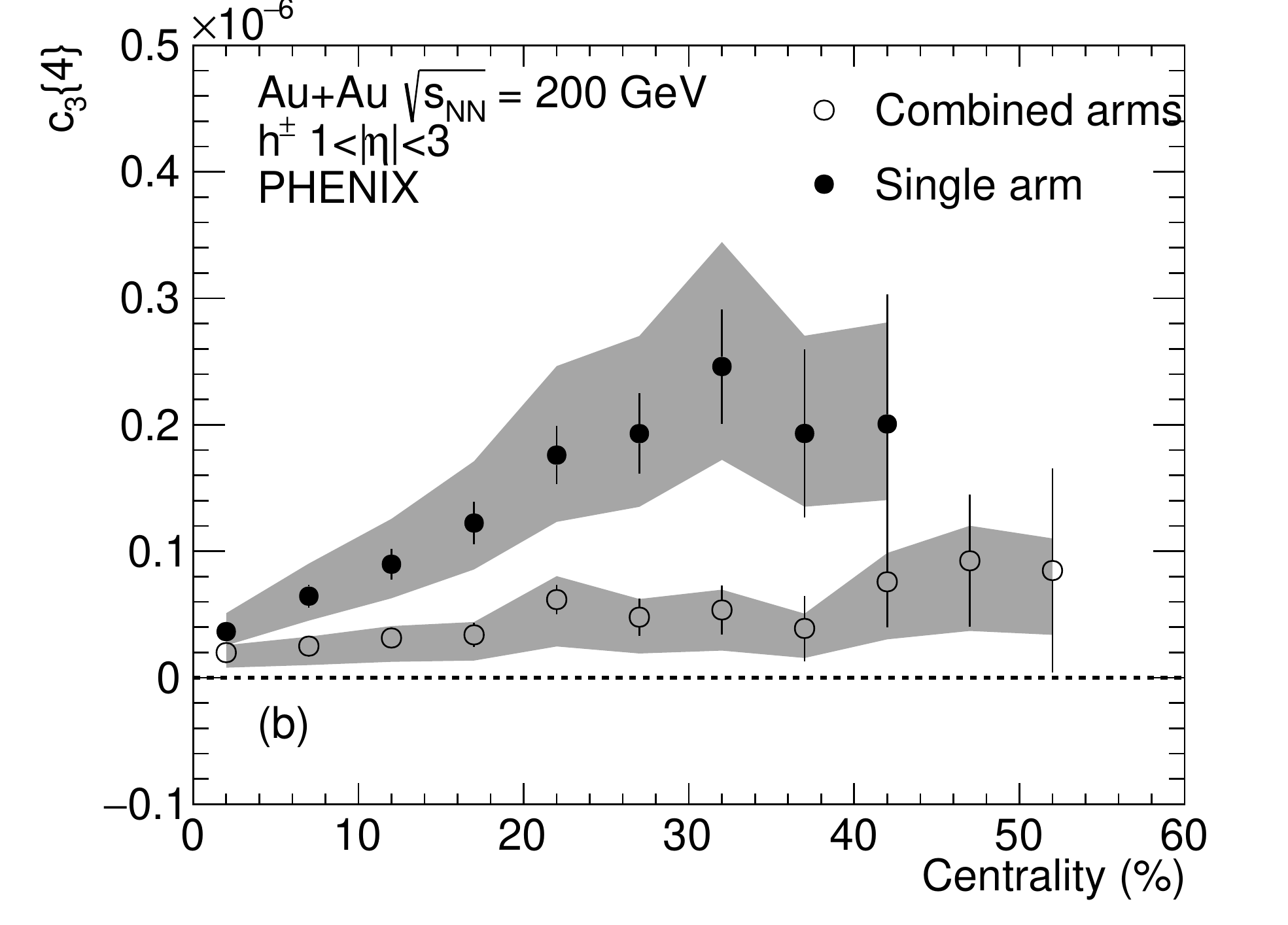}
\caption
{Centrality dependence of $\crf$ for Au$+$Au collisions at
$\snn$ = 200 GeV.
(a) Calculations using both arms: $\crf$ (black circles),
$c_{3}\{4\}_{ab|ab}$ (blue diamonds),
$c_{3}\{4\}_{aa|bb}$ (red squares), and comparison to
STAR~\cite{Adamczyk:2013waa} (black stars).
(b) Comparison of $\crf$ determined using both arms (open symbols)
and a single arm (closed symbols).  Note that the open black circles
are the same in (a) and (b).
}
\label{fig:v34s}
\end{figure*}

Figure~\ref{fig:v34s}~(a) shows the results for $\crf$ as a function of centrality.
The results are always
positive within the systematic uncertainties and shows a trend towards even larger
positive values as one moves away from the most central collisions.  Since
$v_3\{4\} = (-\crf)^{1/4}$ the $v_3\{4\}$ are complex-valued.

Recently positive valued $\ctf$ has been observed in \pau collisions at RHIC~\cite{Aidala:2017ajz}
and \pp collisions at the LHC~\cite{Khachatryan:2016txc,Aaboud:2017blb}
and has been interpreted as arising from short range
nonflow contributions.
The use of subevents, especially when requiring the particles in the cumulant
to be separated in rapidity, significantly reduces nonflow contributions and yields
negative values of $\ctf$ where standard cumulant analysis does not~\cite{Jia:2017hbm}.
In the Au$+$Au analysis presented here, due
to the FVTX acceptance, one includes both short range particle combinations (some or all
particles in a single FVTX arm) and long-range combinations.

We explore the potential influence of such short range nonflow contributions as well as
the opposite effect from long-range decorrelations by changing the FVTX arm requirements
of the particle combinations.  The most extreme is requiring all particles in a single
arm, shown in Figure~\ref{fig:v34s}~(b), and the result is an even larger positive
$\crf$---i.e. in the
direction expected from increased short range nonflow and opposite to the expectation of
long-range decorrelations causing the positive $\crf$.
We can also consider combinations of two subevents, with two particles in each FVTX arm.
One case, labeled ${ab|ab}$, has some short range correlations though fewer than the standard,
whereas the other case, labeled ${aa|bb}$, doesn't allow any.  One sees a consistent behavior emerge:
$\crf_{aa|bb} < \crf_{ab|ab} < \crf < \crf_{\rm single arm}$
All of these results go in the direction of a large nonflow influence
which may be exacerbated by the very small $v_3$ flow signal particularly, at forward
rapidity.

The STAR experiment has also
measured $c_{3}\{4\}$ in Au$+$Au collisions at $\snn$ = 200 GeV, though at
midrapidity $|\eta|<1.0$~\cite{Adamczyk:2013waa}.  Their results, also
shown in Figure~\ref{fig:v34s}~(a), are consistent with zero and fluctuate
between positive and negative $c_{3}\{4\}$ values. The difference
between the STAR and PHENIX data points likely stems from the different
acceptance in pseudorapidity (the STAR points are measured over
$|\eta|<1$ while the PHENIX points are measured over $1<|\eta|<3$ as
discussed above).
Differences in nonflow, event plane decorrelations, and the relative
contribution from fluctuations
as a function of pseudorapidity
may all contribute to these observations.

These results seem to indicate that the small-variance
limit is not applicable to $v_3$ in Au$+$Au collisions at $\snn$ = 200 GeV
for any centrality. Regardless, the measurement of these 2- and 4-particle
cumulants is insufficient to constrain the mean and variance of the
triangular flow event-by-event distribution.

\subsection{Folding Results}
\label{sec:foldingresults}

Now we turn to the results from the event-by-event forward fold. As
detailed in Section~\ref{sec:folding}, in the $v_2$ case the
Bessel-Gaussian parameters are well-constrained apart from the most
central events.  In the $v_3$ case, however, the Bessel-Gaussian
parameters are not well-constrained for any centrality class.  However,
despite the broad range of possible $\delta_{v_3}$ and $v_3^{\rm RP}$
values, these correspond to a rather small range for the real mean
$\mean{v_3}$ and root-mean-square or variance $\sigmar$ of the
distributions.  This means that despite the lack of constraint on the
parameters, the first ($v_3$) and second ($\sigmar$) moments of the
distribution are nevertheless well-constrained.

We can quantify $\mean{v_n}$ and $\sigman$ by varying the
Bessel-Gaussian parameters within the one- and two- standard deviation
statistical constraints.  In addition, we determine the systematic
uncertainties on these quantities by varying the $z$-vertex and
analyzing loose and tight cuts (as described for the cumulants
analysis). An additional systematic uncertainty on the response matrix
is estimated by splitting the data sample into two subsets, one with
higher extracted $\delta$ and one with lower, forward folding the two
data sets separately, and then assessing the difference.

\begin{figure*}[hbtp]
\includegraphics[width=0.99\linewidth]{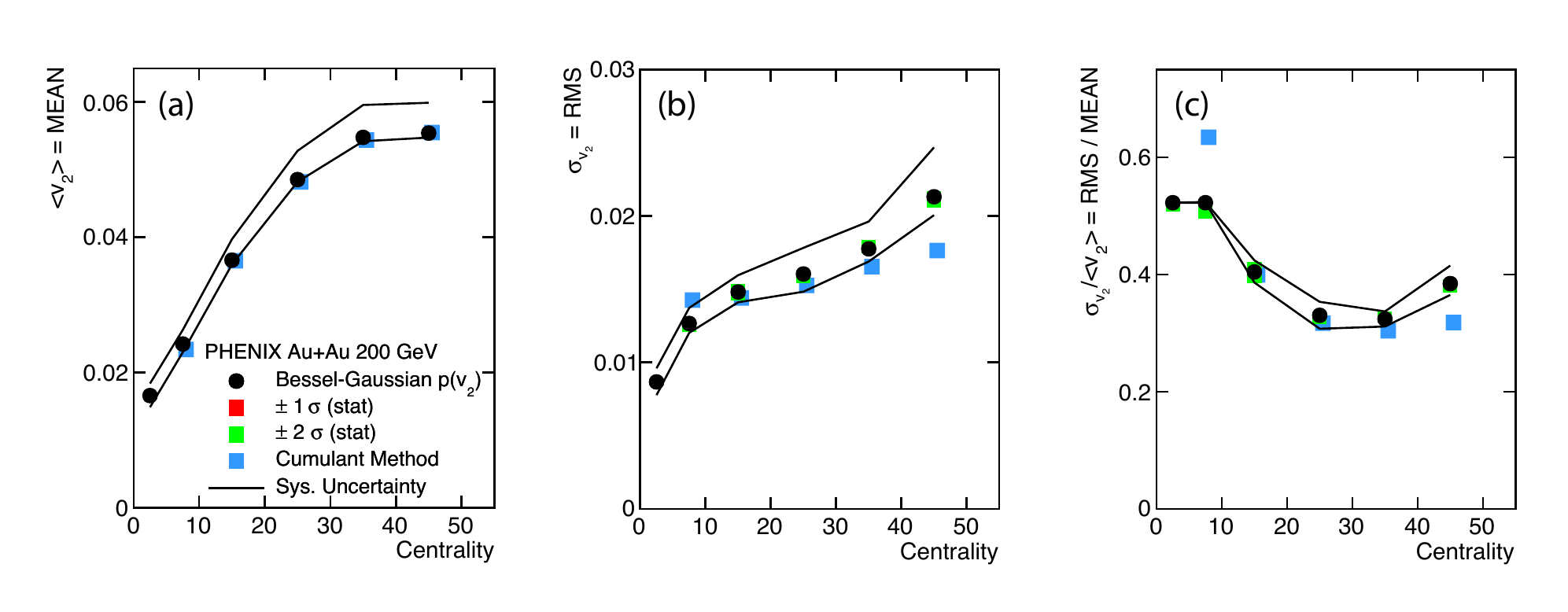}
\caption
{Folding results for (a) $v_2$, (b) $\sigmat$, and (c) $\sigmaot$. The
black lines above and below the points indicate the systematic
uncertainties.  The red (green) boxes indicate the statistical
uncertainties at the 68.27\% (95.45\%) confidence level.  In the case of
$\mean{v_2}$, the statistical uncertainties at the 68.27\% confidence
level are too small to be seen, and the uncertainties at the 95.45\%
confidence level are visible but noticeably smaller than the marker
size. Shown as blue squares are the same quantities as determined using
the cumulant based calculation---these points are slightly offset in the
$x$-coordinate to improve visibility.
}
\label{fig:v2fold}
\end{figure*}

\begin{figure*}[hbtp]
\includegraphics[width=0.99\linewidth]{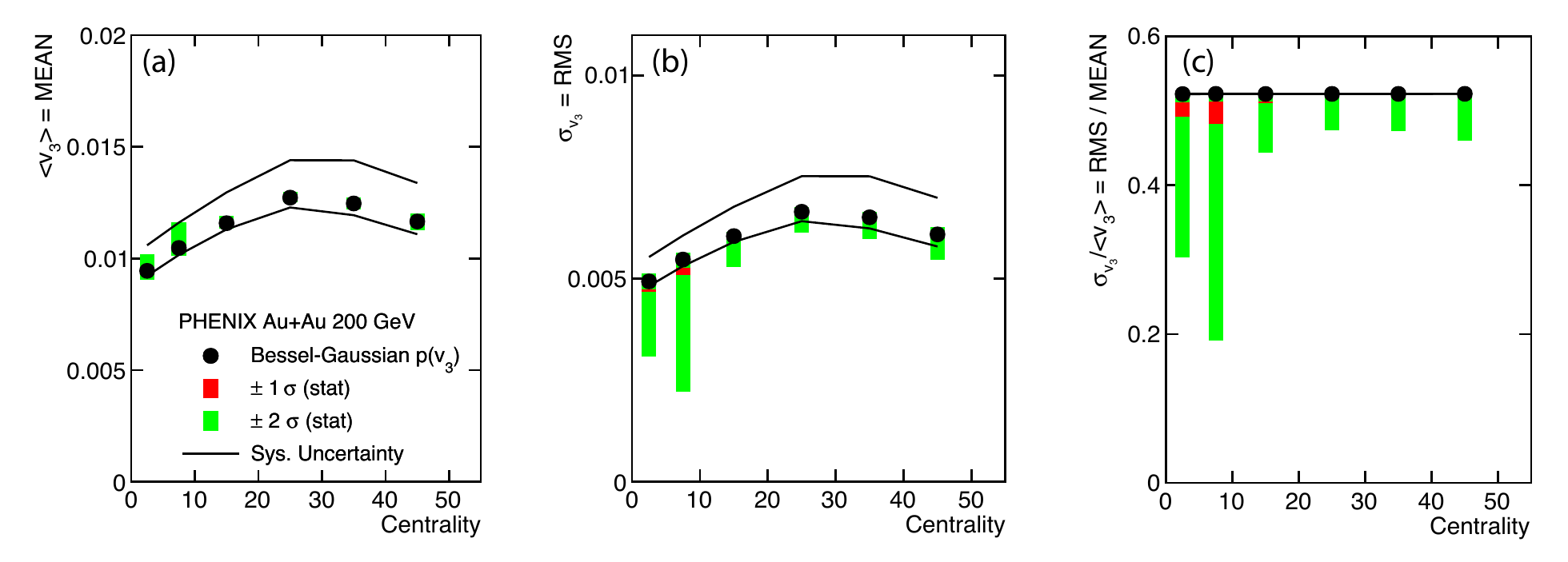}
\caption
{Folding results for (a) $\mean{v_3}$, (b) $\sigmar$, and (c) $\sigmaor$.
The black lines above and below the points indicate the systematic
uncertainties. The red (green) boxes indicate the statistical
uncertainties at the 68.27\% (95.45\%) confidence level.  The $\sigmaor$
values are all $\approx$ 0.52, the apparent limiting value of this
quantity for the Bessel-Gaussian distribution.
}
\label{fig:v3fold}
\end{figure*}

Figure~\ref{fig:v2fold}~(a) shows the extracted first moment
$\mean{v_2}$, Fig.~\ref{fig:v2fold}~(b) shows the extracted second
moment $\sigmat$, and Fig.~\ref{fig:v2fold}~(c) shows the relative
fluctuations $\sigmaot$, each as determined from the folding method and
as a function of centrality. Likewise, Fig.~\ref{fig:v3fold}~(a) shows
the extracted $\mean{v_3}$, Fig.~\ref{fig:v3fold}~(b) shows the
extracted $\sigmar$, and Fig.~\ref{fig:v3fold}~(c) shows the relative
fluctuations $\sigmaor$. The colored bands indicate the statistical
uncertainties at the 68.27\% confidence level (red) and the 95.45\%
confidence level (green) from the $\chi^2$ analysis. The thin black
lines indicate the systematic uncertainties. Also shown
in~\ref{fig:v2fold} as blue squares are results from the cumulant based
calculation as discussed in the previous section.  The $\mean{v_2}$
values are in excellent agreement for all centralities, and the
$\sigmat$ and $\sigmaot$ are in reasonable agreement for 10\%--50\%
centrality, where the small-variance limit holds. Figure~\ref{fig:v32s}
shows a comparison between the cumulant result
$v_3\{2,|\Delta\eta|>2|\}$ and the folding analysis result
$\sqrt{\mean{v_3^2}}$ (calculated from the results in
Fig.~\ref{fig:v3fold}). These results are consistent within the
systematic uncertainties.

We highlight that the $\sigmaot$ values agree well with those determined
from the cumulant method as shown in Figure~\ref{fig:cumufluc}, except
in the most central and peripheral Au$+$Au events.  The most central 0\%--5\% events are
exactly where the Monte Carlo Glauber results in
Figure~\ref{fig:cumufluc} indicate a breakdown in the small-variance
approximation.  This is a good validation of the forward folding
procedure and another confirmation that the event-by-event elliptic flow
fluctuations in Au$+$Au collisions at $\snn$ = 200 GeV are dominated by
initial geometry fluctuations.

\begin{figure}[hbtp]
\includegraphics[width=1.0\linewidth]{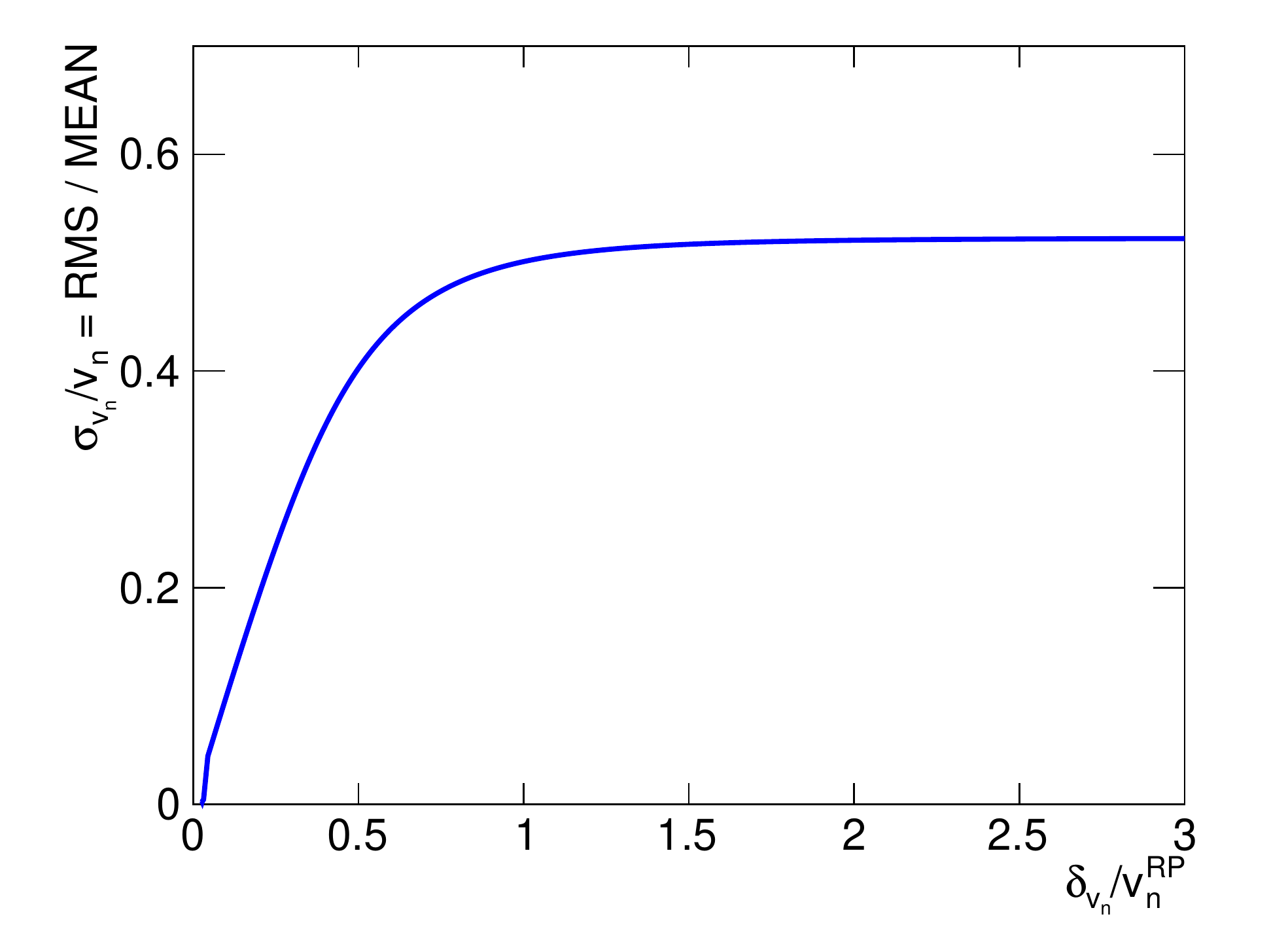}
\caption
{The observed ratio of the mean to the standard deviation (i.e.
$\sigmaon$) of the Bessel-Gaussian as a function of the ratio of the two
free parameters $\delta_{v_n}/v_n^{\rm RP}$.  For values of
$\delta_{v_n} > v_n^{\rm RP}$, the observed $\sigmaon$ saturates at
0.52.
}
\label{fig:limiting}
\end{figure}

Intriguingly, whereas the values of $\sigmaot$ vary significantly as a
function of centrality, the values of $\sigmaor$ are almost precisely
0.52 independent of centrality. To understand this better, we need to
consider a rather peculiar feature of the Bessel-Gaussian Function.
Figure~\ref{fig:limiting} shows the $\sigmaon$ of the Bessel-Gaussian as
a function of the ratio $\delta/v_n^{\rm RP}$.  For values of $\delta >
v_n^{\rm RP}$, the observed $\sigmaon$ saturates at a value of about
0.52.  Thus, any Bessel-Gaussian in the large variance limit will have a
$\sigmaon$ of the same value.

\begin{figure}[hbtp]
\includegraphics[width=1.0\linewidth]{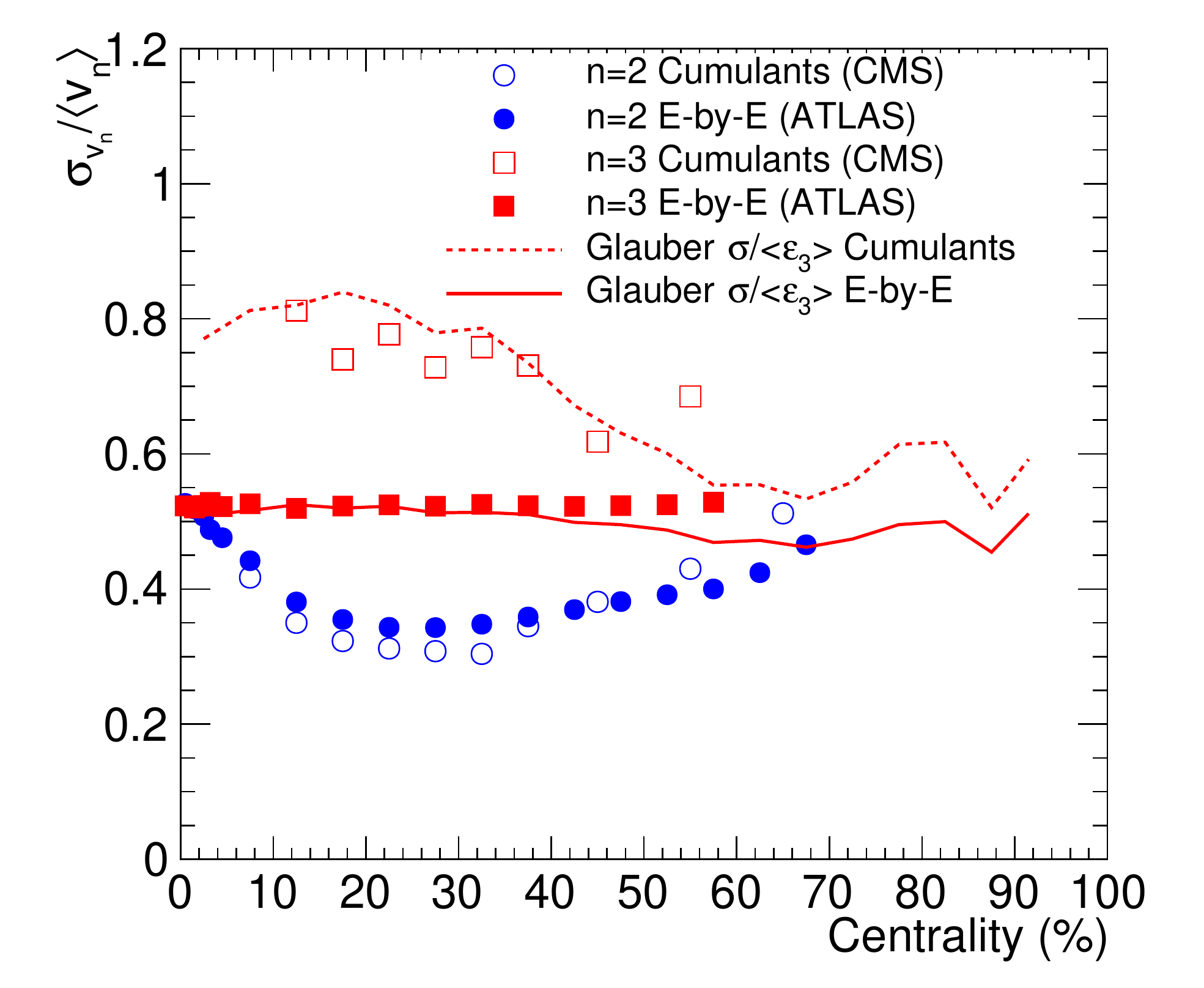}
\caption
{The observed ratio of the standard deviation to the mean (i.e.
$\sigmaon$) for $n=2$ and $n=3$ as a function of centrality in Pb$+$Pb
collisions at $\snn$~=~2.76~TeV~\cite{Chatrchyan:2013kba,Aad:2013xma}.
Also shown are $\sigma_{\eps_3}/\eps_3$ values from MC Glauber estimated
using the small-variance limit cumulant technique (shown as the dashed
line) and the direct calculation from the moments (shown as the solid
line).
}
\label{fig:lhc}
\end{figure}

This observation can, in fact, help shed light on the observed
discrepancy between the CMS~\cite{Chatrchyan:2013kba} and
ATLAS~\cite{Aad:2013xma} data on $\sigmaor$. Figure~\ref{fig:lhc} shows
$\sigmaot$ and $\sigmaor$ as a function of centrality in Pb$+$Pb
collisions at $\snn$ = 2.76 TeV from CMS and ATLAS.  The CMS results are
obtained using the cumulant method assuming the small-variance limit. In
contrast the ATLAS results are obtained via an event-by-event unfolding
and calculating the exact mean and variance of the distribution.

The $\sigmaot$ values are in very good agreement, which appears to
validate the small variance approximation (as was also validated in the
Au$+$Au at $\snn$ = 200 GeV case in this analysis). In contrast, there
is a large difference in the $\sigmaor$ between the different methods.
The ATLAS $\sigmaor$ values are all very close to 0.52, exactly as
observed above in the present Au$+$Au data and as found to be a limiting
case for the Bessel-Gaussian function.  To better understand the
$\sigmaor$, we also show $\sigma_{\eps_3}/\mean{\eps_3}$ as determined
from MC Glauber calculations. The dashed red-line uses the
small-variance limit estimate with cumulants, as is done for the CMS
data, and the agreement is quite reasonable.  The solid red line is
calculated from the moments of the $\eps_3$ distribution directly, and
shows good agreement with the ATLAS data.  This represents a
quantitative confirmation of the event-by-event fluctuations and the
breakdown in the small variance approximation.
The $v_3\{4\}$ at forward rapidity at RHIC is found to be complex-valued,
which may be the result of a very small flow $v_3$ and significant nonflow contributions.

\section{Summary and conclusions}
\label{sec:conclusions}

In summary, we have presented measurements of elliptic and triangular
flow in \auau collisions at 200 GeV for charged hadrons at forward
rapidity $1 < |\eta| < 3$.  In particular, we compare flow cumulants
($\vtt$, $\vtf$, $\vts$, $\vte$ and $v_3\{2\}$, $v_3\{4\}$) and the mean
and variance of the $v_2$ and $v_3$ event-by-event distributions using a
forward-fold procedure with a Bessel-Gaussian ansatz.  These
measurements are complementary in terms of sensitivity to initial state
geometry fluctuations and additional fluctuations from the evolution of
the medium, for example via dissipative hydrodynamics.

In the small-variance limit, where the event-by-event flow fluctuations
are small compared to the average flow value i.e. $\sigmaon < 1$, we
expect the cumulants extraction and the forward-fold results to agree.
This is the case for elliptic flow in \auau collisions from 10\%--50\%
central and both results agree with event-by-event fluctuations in the
initial geometry as calculated via Monte Carlo Glauber.

In contrast, we find that the small-variance limit fails for triangular
flow for all centralities at RHIC and the LHC. For LHC \pbpb results,
the large-variance result for the cumulants can be described purely via
Monte Carlo Glauber initial geometry fluctuations.  However, for RHIC
\auau collisions the complex values of $v_{3}\{4\}$ indicate that there
may be additional
nonflow influences as well as
sources of fluctuations in the translation of initial
geometry into final state momentum triangular anisotropies.  Detailed
comparisons with event-by-event hydrodynamic calculations should be
elucidating to understand the nature of these fluctuations.


\section*{ACKNOWLEDGMENTS}   

We thank the staff of the Collider-Accelerator and Physics
Departments at Brookhaven National Laboratory and the staff of
the other PHENIX participating institutions for their vital
contributions.  We acknowledge support from the
Office of Nuclear Physics in the
Office of Science of the Department of Energy,
the National Science Foundation,
Abilene Christian University Research Council,
Research Foundation of SUNY, and
Dean of the College of Arts and Sciences, Vanderbilt University
(U.S.A),
Ministry of Education, Culture, Sports, Science, and Technology
and the Japan Society for the Promotion of Science (Japan),
Conselho Nacional de Desenvolvimento Cient\'{\i}fico e
Tecnol{\'o}gico and Funda\c c{\~a}o de Amparo {\`a} Pesquisa do
Estado de S{\~a}o Paulo (Brazil),
Natural Science Foundation of China (People's Republic of China),
Croatian Science Foundation and
Ministry of Science and Education (Croatia),
Ministry of Education, Youth and Sports (Czech Republic),
Centre National de la Recherche Scientifique, Commissariat
{\`a} l'{\'E}nergie Atomique, and Institut National de Physique
Nucl{\'e}aire et de Physique des Particules (France),
Bundesministerium f\"ur Bildung und Forschung, Deutscher
Akademischer Austausch Dienst, and Alexander von Humboldt Stiftung (Germany),
J. Bolyai Research Scholarship, EFOP, the New National Excellence
Program ({\'U}NKP), NKFIH, and OTKA (Hungary),
Department of Atomic Energy and Department of Science and Technology (India),
Israel Science Foundation (Israel),
Basic Science Research and SRC(CENuM) Programs through NRF funded by the 
Ministry of Education and the Ministry of Science and ICT (Korea),
Physics Department, Lahore University of Management Sciences (Pakistan),
Ministry of Education and Science, Russian Academy of Sciences,
Federal Agency of Atomic Energy (Russia),
VR and Wallenberg Foundation (Sweden),
the U.S. Civilian Research and Development Foundation for the
Independent States of the Former Soviet Union,
the Hungarian American Enterprise Scholarship Fund,
the US-Hungarian Fulbright Foundation,
and the US-Israel Binational Science Foundation.


\section*{APPENDIX:  Test case for full unfold}
\label{APPENDIX}

\begin{figure*}[th]
\includegraphics[width=0.8\linewidth]{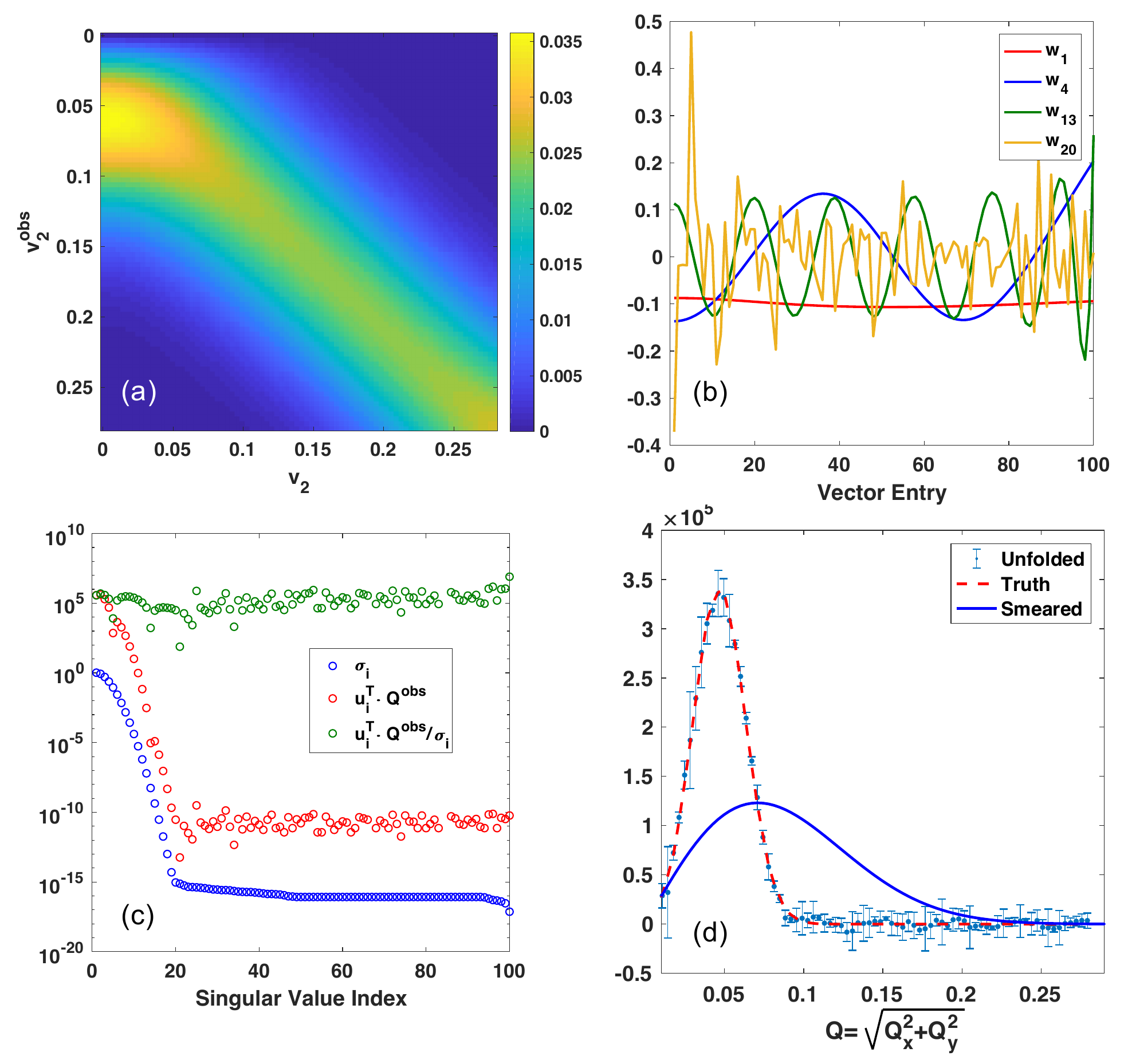}
\caption{(a) Response matrix for test unfolding problem; (b) Selected
right-singular values of the response matrix; (c) Picard plot for
inverse problem $\hat{A}\vec{Q}_2^{{\rm true}}=\vec{Q}_2^{\rm obs}$,
see text for details; (d) True, smeared, and unfolded $Q_2$ as
determined using SVD with attenuation factors.
}
\label{fig_toy_svd}

\includegraphics[width=0.99\linewidth]{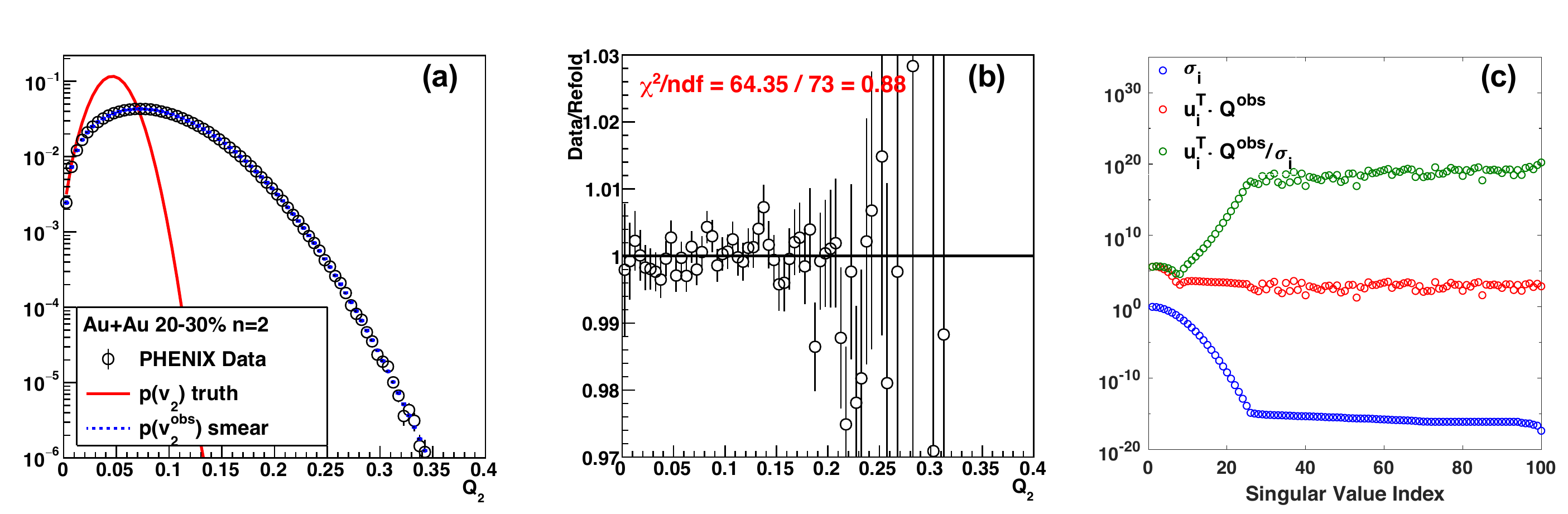}
\caption{(a) Data distribution for $Q_{2}^{\rm meas}$ in Au$+$Au collisions
at $\sqrt{s_{_{NN}}}$ = 200 GeV in the
20\%--30\% centrality class.  Also shown is an assumed Bessel-Gaussian truth
distribution and its resolution smeared result. (b) Data divided by the
resolution smeared solution showing a good agreement within statistical
uncertainties. (c) Picard plot for inverse problem with data
$\hat{A}\vec{Q}_2^{{\rm true}}=\vec{Q}_2^{\rm obs}$, see text for
details;
}
\label{fig_data_picard_unfold}
\end{figure*}

For this test case, the response matrix $\hat{A}$ is shown in
Fig.~\ref{fig_toy_svd}~(a) and is identical to that for the real data
20\%--30\% centrality class.  We then attempt to solve the inverse problem
$\hat{A}\vec{Q}_2^{{\rm true}}=\vec{Q}_2^{{\rm obs}}$, where
$Q_2^{{\rm obs}}$ has been obtained in the limit of infinite
statistical precision, assuming a truth-level distribution with
parameters such that the smearing, as encoded in $\hat{A}$, yields a
distribution similar to that measured in data. The singular value
factorization $\hat{A}=\hat{U}\hat{\Sigma}\hat{W}^{T}$ of the matrix is
obtained, where $\hat{U}$ and $\hat{W}$ are unitary matrices whose
column vectors, $u_i$ and $w_i$, are the left- and right- singular
vectors of $\hat{A}$, respectively, and $\hat{\Sigma}$ is a diagonal
matrix, whose nonzero entries $\sigma_i$ are its singular values.
Figure~\ref{fig_toy_svd}~(b) shows a few selected right-singular vectors
$w_i$. Notice that some vectors, namely those corresponding to the
largest singular values, are harmonic, whereas those corresponding to
the smallest singular values are essentially noise.

Because the response matrix is singular, we use the SVD decomposition to
construct the solution of the inverse problem as a linear combination of
\textit{all} right-singular vectors, as follows:
\begin{equation}
\label{eq_svd}
\vec{Q}_{2} = \sum_{i=1}^{{\rm Dim}(A)}\varphi_i\left ( \frac{\vec{u}_i^{T}\cdot
  \vec{Q}_2^{{\rm obs}}}{\sigma_i} \right )\vec{w}_i.
\end{equation}
The damping factors $\varphi_i = \sigma_{i}^2/(\sigma_{i}^2 +
\lambda^2)$, for some $\lambda\in\mathbb{R}$, are introduced to
attenuate the contribution of the noisy singular vectors to the sum. It
is important to point out that in most implementations of SVD used in
high-energy physics, including RooUnfold, the above sum is simply
truncated to include only a subset of the harmonic singular vectors,
potentially leading to loss of information.

To determine which singular vectors contribute to the solution
in a meaningful manner, it is useful to examine the \textit{Picard
plot}~\cite{Hansen1990} for the problem at hand, shown in
Fig.~\ref{fig_toy_svd}~(c), which displays the singular values
$\sigma_i$ of $\hat{A}$, as well as the projection of $\vec{Q_2^{\rm
obs}}$ onto the singular vectors $\vec{u}_i^{T}\cdot \vec{Q}_2^{\rm
obs}$, and the solution coefficients $\vec{u}_i^{T}\cdot \vec{Q}_2^{\rm
obs}/\sigma_i$. Notice that the singular values and the Fourier
coefficients drop sharply many orders of magnitude before leveling off,
yet in such a way that their ratio is roughly constant. The implication
is then that all singular vectors appear to contribute equally to the
solution, which is clearly problematic given the noisy nature of most of
them. In general, it is desirable for Fourier coefficients to drop off
faster than the singular values (to fulfill the so-called discrete
Picard condition), such that the Picard plot will reveal the appropriate
set of terms to include in the solution, as identified by a sharp drop
in the solution coefficients.

Given that our problem does not satisfy the Picard condition, we
introduce the attenuation factors $\varphi_i$ in Eqn.~\ref{eq_svd}. The
resulting unfolded $Q_2$ is shown in Fig.~\ref{fig_toy_svd}~(d), along
with the true $Q_2^{{\rm true}}$, and smeared $Q_2^{\rm obs}$. We
observe that the unfolding works well, yielding a good description of
the true distribution shape, with uncertainties associated with varying
the regularization parameter $\lambda$.

However, in this case the unfolding procedure constitutes an ill-posed
inverse problem, such that small perturbations in the input
vector---that is, $Q_2^{\rm obs}$---translate to very large errors in
the solution, compounded by the fact that the Picard condition is
violated. In particular, we have verified with our test problem that the
statistical fluctuations in $Q_2^{\rm obs}$ when sampling a
\textit{finite} number of events, comparable to those recorded in data,
indeed limit the number of available harmonic singular vectors, thus
causing the solution to be dominated by noise.

We now examine the application of the above unfolding method to data.
Fig.~\ref{fig_data_picard_unfold}~(a) shows an \textit{ansatz} for
$Q_2^{{\rm true}}$ assuming a Bessel-Gaussian form, and the
corresponding refolded smeared distribution. It compares very well to
the data, as shown in the ratio plot in
Fig.~\ref{fig_data_picard_unfold}~(b). In principle, given the good
quality of the fit, one would expect the unfolding procedure to work
with the data as input. However, the statistical fluctuations apparent
in the ratio plot perturb the solution in such a way that the noisy
nonharmonic singular vectors are enhanced even more than in the test
problem, as shown in Fig.~\ref{fig_data_picard_unfold}~(c). As a result,
the number of available harmonic singular vectors is reduced, and the
problem has no satisfactory solution, even when regularization is
applied.  Thus, to be explicit, the unfolding procedure fails.  We note
that if we apply our test example with a significantly better
resolution, i.e. as in the ATLAS Pb$+$Pb case, the method does converge as
expected.

%
 
\end{document}